\documentclass[lettersize,journal]{IEEEtran}
\usepackage{amsmath,amsfonts}
\usepackage{algorithmic}
\usepackage{algorithm}
\usepackage{array}
\usepackage[caption=false,font=normalsize,labelfont=sf,textfont=sf]{subfig}
\usepackage{textcomp}
\usepackage{stfloats}
\usepackage{url}
\usepackage{verbatim}
\usepackage{graphicx}
\usepackage{cite}
\usepackage{multirow}
\usepackage{threeparttable}
\usepackage{booktabs}
\usepackage{subfig} 
\usepackage{subfloat}
\usepackage[normalem]{ulem}
\useunder{\uline}{\ul}{}
\usepackage{longtable}\usepackage{array}
\usepackage{stackengine}

\usepackage{bbding}
\usepackage{braket}
\usepackage{qcircuit}
\usepackage{makecell}  
\usepackage{xcolor}  
\usepackage{tablefootnote}
\usepackage[T1]{fontenc}
\usepackage{booktabs}

\begin{document}
	
	\title{QGHNN: A quantum graph Hamiltonian \\ neural network}
	
	\author{Wenxuan Wang,~\IEEEmembership{Student Member,~IEEE,} 
		\thanks{Wenxuan Wang is with the School of Computer Science and Engineering, Central South Univerisity, Changsha, China, 410083, with the Department of Computer Science, City University of Hong Kong, Hong Kong, 999077 and with the Key Lab of MIMS, College of Computer Science and Technology, Guangxi Normal University, Guilin, China, 541004. (Email: 234701002@csu.edu.cn).}
	}
	
	\markboth{Journal of \LaTeX\ Class Files,~Vol.~14, No.~8, August~2021}%
	{Shell \MakeLowercase{\textit{et al.}}: A Sample Article Using IEEEtran.cls for IEEE Journals}
	
	
	\maketitle
	
	\begin{abstract}
		Representing and learning from graphs is essential for developing effective machine learning models tailored to non-Euclidean data. 
		While Graph Neural Networks (GNNs) strive to address the challenges posed by complex, high-dimensional graph data, Quantum Neural Networks (QNNs) present a compelling alternative due to their potential for quantum parallelism. 
		However, much of the current QNN research tends to overlook the vital connection between quantum state encoding and graph structures, which limits the full exploitation of quantum computational advantages. 
		To address these challenges, this paper introduces a quantum graph Hamiltonian neural network (QGHNN) to enhance graph representation and learning on noisy intermediate-scale quantum computers.
		Concretely, a quantum graph Hamiltonian learning method (QGHL) is first created by mapping graphs to the Hamiltonian of the topological quantum system.
		Then, QGHNN based on QGHL is presented, which trains parameters by minimizing the loss function and uses the gradient descent method to learn the graph.
		Experiments on the PennyLane quantum platform reveal that QGHNN outperforms all assessment metrics, achieving the lowest mean squared error of \textbf{$0.004$} and the maximum cosine similarity of \textbf{$99.8\%$}, which shows that QGHNN not only excels in representing and learning graph information, but it also has high robustness ability. 
		QGHNN can reduce the impact of quantum noise and has significant potential application in future research of quantum knowledge graphs and recommendation systems.
	\end{abstract}
	
	
	\section{Introduction}
	\IEEEPARstart{Q}{uantum} machine learning (QML) integrates the advantages of quantum computing and machine learning, aiming to use fundamental principles of quantum mechanics to enhance the performance and efficiency of classical algorithms, which has gained extensive interest\cite{ ref.tetci.Dixit, ref.tetci.Liu}.
	In 2022, Google proved the quantum advantages of QML models in learning tasks, including predicting observable and performing quantum principal component analysis on noisy intermediate-scale quantum computers (NISQ) \cite{ref.huang2022quantum}. 
	With the development and invention of NISQ devices to 1121 qubits, the performance and computational efficiency of QML models have been greatly enhanced. 
	For example, Shi et al. developed a pretrained quantum-inspired deep neural network in 2024, to achieve high performance and interpretability in relevant NLP disciplines \cite{ref.tcyb.04.shi}. 
	Yu et al. introduced a simple and general reward design method based on quantum machine learning to effectively simulate quantum systems of different dimensions \cite{ref.tetci.Yu}. 
	Zheng et al. developed a QML model for graph classification based on spatial graph convolutional neural networks, improving the ability of quantum neural networks to handle non-Euclidean data problems \cite{ref.zheng2024quantum}.
	Graph, a typical representation of non-Euclidean data\cite{ref.zhang.2022graph}, has various applications in object interaction\cite{ref.tetci.LiuGuanfeng}, recommendation systems\cite{ref.tetci.Zhu}, and natural language processing\cite{ref.zhang.wang2021branch}.
	Quantum Neural Networks (QNNs) apply the characteristics of quantum superposition and entanglement to process multiple states simultaneously \cite{ref.zhao2024gqhan}, which allows QNNs to effectively compute the connections between different nodes and edges, increasing the computational efficiency of QML models \cite{ref.QML.biamonte2017quantum}.
	In recent years, there has been much interest in constructing QNN models to represent and learn graphs.
	Bai et al. created a quantum graph convolutional neural network based on continuous-time quantum walks and average mixing matrices to extract multi-scale node properties from graphs \cite{ref.bai2021learning}. 
	Zhang et al. developed a quantum subgraph neural network that efficiently handles graph classification tasks by capturing global and local graph structures \cite{ref.zhang2019quantum}. 
	Dernbach et al. created a graph neural network using quantum random walks to calculate diffusion operators \cite{ref.dernbach2019quantum}.
	A series of QNN models have been presented, encouraging the research process of learning graphs in QML.
	However, the potential of using QNN to learn graphs on NISQ devices has not been thoroughly exploited. 
	Generally, the QNNs mentioned above are primarily quantum-inspired QML models that do not include specific quantum circuit construction methodologies, which limits the ability of QNN to represent and learn graphs on NISQ devices and impedes the future development of QML.
	As a result, developing an appropriate quantum graph neural network model to represent and learn graphs has become a critical task.

	Fortunately, we discover that quantum Hamiltonian learning (QHL), a method that employs parameterized quantum circuits to represent and learn the Hamiltonian of the quantum system, has the ability to complete this task \cite{ref.Wiebe2014HL,ref.hamiltonian.2024.NC}. 
	QHL builds parameterized quantum circuits based on creating and decomposing unitary operators with Hamiltonian and parameters, speeding up the development of QML models for learning quantum systems. 
	Several QHL algorithms have been employed to perform application tasks on NISQ devices, like image segmentation \cite{ref.shi2022parameterized.NEW}, generating models \cite{ref.araz2023quantum.HL}, and predicting observables \cite{ref.koch2023adversarial.HL}. 
	QHL can represent and learn information from specific quantum systems, and it can be applied to calculating high-dimensional data as quantum computers advance.
	This research aims to establish the correlation between graph information and the Hamiltonian of a quantum system by designing quantum circuits based on QHL and using NISQ quantum devices to represent and learn graph information effectively. 
	
	Quantum graph Hamiltonian learning (QGHL) is presented as a method of mapping graph to the Hamiltonian of a topological quantum system, which applies the lattice characteristics of the topological quantum system to develop parameterized quantum circuits and produces a final quantum system state containing graph information.
	Quantum graph Hamiltonian neural network (QGHNN) based on QGHL is developed to train the parameters of quantum circuits by minimizing the loss function and using gradient descent methods to represent and learn graphs on NISQ devices. 
	Furthermore, experimental results on the PennyLane quantum platform show that QGHNN is not only highly efficient at representing and learning graphs but also has high robustness ability, which can reduce the influence of quantum noise and suggests that QGHNN may have various potential applications in high-dimensional data processing.

	The main contributions made by this work are outlined in the following lines.
	\begin{itemize}
		\item QGHL is proposed, establishing a mapping link between graph and the Hamiltonian of topological quantum systems, which provides considerable advantages in representing and learning target graph.  
		
		\item QGHNN is presented, which updates the parameters of quantum circuits by minimizing the loss function and employing gradient descent methods, indicating exciting applications in graph analysis on NISQ devices.  
		
		\item The results of experiments conducted on the PennyLane quantum platform show that QGHNN can not only represent and learn classical graph information but also has high robustness ability, which reduces the influence of quantum noise. QGHNN outperforms all assessment metrics, with the lowest mean squared error ($0.004$) and the highest cosine similarity ($99.8\%$). 
		
	\end{itemize}
	
	The structure of the subsequent sections is as follows. 
	Section 2 overviews the basic principles of QML and other related work. 
	Section 3 describes QGHL. 
	Section 4 introduces QGHNN based on QGHL. 
	Section 5 presents the results and discussion of the experiment. 
	The conclusion is shown in Section 6.

	\section{Related work}
	
	\label{section_ii}
	
	This section starts with a review of the foundations in QML, followed by an overview of the GNN and QHL method. 
	Table. (\ref{tab:notations}) summarizes the notations.
	\begin{table}[htpb]
		\renewcommand{\arraystretch}{1.6}
		\caption{Notations.}  
		\label{tab:notations}
		\centering
		\begin{tabular}{p{3.7cm}<{\centering} p{3.7cm}<{\centering}}  
			\hline  
			Notation & Description\\   \hline
			$H_m$, $H_c$, $H_t$, $H$ & Hamiltonian\\
			$J$ &  Coupling strength \\
			$U(\theta)$ &  Unitary operator \\
			$|\psi_{in}\rangle$, $|\psi_{out}\rangle$, $|\psi_{t}\rangle$ & Quantum state \\
			$R_{x}(\theta)$, $R_{y}(\theta)$, $R_{z}(\theta)$ & Rotation gate \\
			$\sigma^{x}$, $\sigma^{y}$, $\sigma^{z}$ &  Pauli operator  \\ \hline  
		\end{tabular}  
	\end{table} 	
	\subsection{Preliminary}
	The foundation of QGHNN in quantum machine learning is explained as follows.
	
	\setlength{\parindent}{0.35cm}
	\textbf{Quantum state:} 
	Quantum state represents the state of a qubit, which is the fundamental unit of quantum computing.
	Mathematically, the quantum state $|\psi\rangle$ of a qubit can be represented as a two-dimensional complex vector in Eq.(\ref{Eq:Quantum state}), 
	\begin{equation}
		\label{Eq:Quantum state}
		|\psi\rangle = \alpha|0\rangle + \beta|1\rangle,
	\end{equation}
	where $\alpha$ and $\beta$ are complex numbers, satisfying the normalization condition $|\alpha|^2 + |\beta|^2 = 1$. 
	The quantum state changes over time, which is referred to as the evolution of the quantum system. 
	
	\setlength{\parindent}{0.35cm}
	\textbf{Quantum gate:} 
	Quantum gates are the fundamental operational elements of quantum computing and are often employed in QML models, which can be divided into unitary operator $U(\theta)$ in Eq.(\ref{Eq:Unitary operators}) \cite{ref.quantum.gate.unitory.2021},
	\begin{equation}
		\label{Eq:Unitary operators}
		U(\theta) = e^{-\frac{iH\theta}{\hbar}},
	\end{equation}
	where $\hbar$ is the Planck constant, $H$ represents the Hamiltonian, and $i$ is related to the imaginary.
	The evolution of the quantum state $|\psi(\theta)\rangle$ can be represented by the unitary operator $U(\theta)$ in Eq.(\ref{Eq:state evolution}),
	\begin{equation}
		\label{Eq:state evolution}
		|\psi(\theta)\rangle = U(\theta)|\psi_0\rangle.
	\end{equation}
	Table. (\ref{tab: quantum gate}) summarizes the quantum gates that are used in parameterized quantum circuits of QGHNN.
	
	\setlength{\parindent}{0.35cm}
	\textbf{Hamiltonian:} 
	Hamiltonian is important in determining a quantum system and its evolution \cite{ref.Wiebe2014HL}, which is related to the energy of the whole quantum system.
	The unitary operator based on Hamiltonian controls the system evolution of quantum state.
	
	\setlength{\parindent}{0.35cm}
	\textbf{Topological Quantum System:}
	Topological quantum system is characterized by lattice and topological properties, which means that its properties remain constant in the face of some quantum noise influence \cite{ref.QHL.2023}.
	Hamiltonian is an essential part of describing the topological system and its evolution.

    \begin{table*}[ht]
		\setlength{\abovecaptionskip}{0cm}  
		\setlength{\belowcaptionskip}{-0.2cm} 
		\renewcommand\arraystretch{1.8}
		\centering
		\caption{Quantum gates}
		\setlength{\tabcolsep}{11.3mm}{
			\begin{tabular}{lccc}
				\toprule
				Name & Symbol & Matrix Reception & Symbol \\
				\midrule
				Rotation Pauli X gate   &  $R_{x}(\theta)$ Gate
			&  $\left[\begin{array}{cc}\cos \frac{\theta}{2} & -i\sin \frac{\theta}{2} \\ -i\sin \frac{\theta}{2} & \cos \frac{\theta}{2}\end{array}\right]$ & \begin{minipage}[b]{0.23\columnwidth}
				\centering
				\raisebox{-.5\height}{\includegraphics[width=\linewidth]{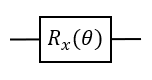}}
			\end{minipage} \\
			\specialrule{0em}{0em}{10pt}
				Rotation Pauli Y gate    & $R_y{(\theta)}$ & $\left[\begin{array}{cc}\cos \frac{\theta}{2} & -\sin \frac{\theta}{2} \\ \sin \frac{\theta}{2} & \cos \frac{\theta}{2}\end{array}\right]$ & \begin{minipage}[b]{0.23\columnwidth}
				\centering
				\raisebox{-.5\height}{\includegraphics[width=\linewidth]{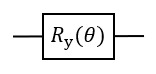}}
			\end{minipage} \\
			\specialrule{0em}{0em}{10pt}
				Rotation Pauli Z gate     & $R_z{(\theta)}$ & $\left[\begin{array}{cc}e^{-i \frac{\theta}{2}} & 0 \\ 0 & e^{i \frac{\theta}{2}}\end{array}\right]$ & \begin{minipage}[b]{0.22\columnwidth}
				\centering
				\raisebox{-.5\height}{\includegraphics[width=\linewidth]{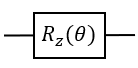}}
			\end{minipage} \\
			\specialrule{0em}{0em}{10pt}
				CNOT gate & $CNOT$ &$\left[\begin{array}{llll}1 & 0 & 0 & 0 \\ 0 & 1 & 0 & 0 \\ 0 & 0 & 0 & 1 \\ 0 & 0 & 1 & 0\end{array}\right]$  & \begin{minipage}[b]{0.23\columnwidth}
				\centering
				\raisebox{-.5\height}{\includegraphics[width=\linewidth]{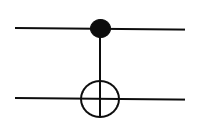}}
			\end{minipage}  \\
				\bottomrule
			\end{tabular}
		}
		\label{tab: quantum gate}
	\end{table*}
	
	\subsection{Graph Neural Network}
	
	\begin{figure}[t]
		\centering
		\includegraphics[width=0.48\textwidth]{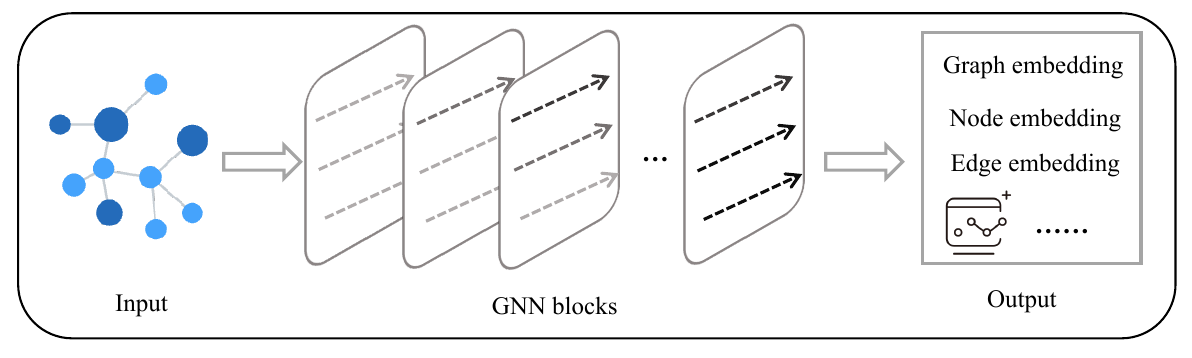}
		\caption{Graph Neural Network. The input graph data can be trained by GNN to realize the data representation.}
		\label{fig:Graph neural network}
	\end{figure}
	
	Graph Neural Networks (GNNs) are deep learning models that are specifically designed to process graph-structured data \cite{ref.zhang.peng2022relation}. 
	They are frequently used in a variety of applications, including social network analysis, recommendation systems, image segmentation, and bioinformatics \cite{ref.tetci.graph01, ref.tetci.graph02}. 
	Nodes and edges are the basic building blocks of a graph, with nodes representing entities and edges indicating their relationships. 
	GNN uses a message-passing method to effectively capture nodes interactions and the global structural information of the graph, increasing the efficacy of feature learning. 
	
	Fig. (\ref{fig:Graph neural network}) depicts the framework of a GNN, beginning on the left, the input portion displays a complicated graph structure, demonstrating the interactions between various nodes and edges \cite{ref.tcyb.21.graph.LXL}. 
	Nodes are depicted in various hues of blue, offering a clear visual picture of their connections and the structural complexity of the network. 
	The feature vectors of each node provide rich information for later processing, setting the groundwork for the feature learning of GNN and allowing it to efficiently extract useful information from graph data. 
	The collection of GNN blocks that represent the flow of information and the dynamic updates of node features via arrows. During the feature aggregation process, each GNN block takes into account data from surrounding nodes, gradually optimizing node representations to capture more complex graph structural aspects. 
	The final output section shows the created graph embeddings, node embeddings, and edge embeddings, demonstrating the wide range of feature representations that GNN can produce while processing graph data. 
	These results can be used for a variety of downstream tasks, including classification, clustering, and recommendation systems, demonstrating the broad capabilities of GNN in processing graph data with complex interactions \cite{ref.zhang.liu2022learning,ref.tetic.LXL.zhang2020reconstructing}. 
	Despite its outstanding performance in a wide range of applications, GNN confronts a number of hurdles \cite{ref.tetic.zhang.peng2022relation}. 
	The development of quantum machine learning provides potential new approaches to addressing these difficulties. 
	GNN can process complicated graph more efficiently by leveraging the parallelism of quantum machine learning and the framework of QHL. 
	QHL uses quantum states and Hamiltonian to better capture the interactions between nodes and edges in a graph, hence increasing the expressiveness and predictive accuracy of QML models. 
	
	\subsection{Quantum Hamiltonian Learning}
	
	QHL is a quantum machine learning method that effectively acquires information about the Hamiltonian of a quantum system by training parameterized quantum circuits \cite{ref.shi2022parameterized.NEW}. 
	The primary objective of QHL is to precisely represent and learn the Hamiltonian $H_{target}$ of quantum systems by developing and creating unitary operators $U(\theta)$ with adjustable parameters. 
	QHL allows the initial quantum system $|\psi_{in}\rangle$ to generate the final quantum system $|\psi_{out}\rangle$ that correctly simulates the target Hamiltonian $H_{target}$.
	QHL comprises four primary components: the initial quantum system $|\psi_{in}\rangle$, the quantum system evolution executed by the parameterized quantum circuits, the optimizer responsible for defining the loss function, and the final quantum system $|\psi_{out}\rangle$. 
	In QHL, the initial quantum system $|\psi_{in}\rangle$ experiences system evolution through unitary operators $U(\theta)$ that are constructed by using parameterized quantum circuits, and $\theta$ represents the different kinds of parameters that need to be optimized.
        The parameters in the same class of quantum gates are the same.
	The unitary operator $U(\theta)$ can be built with the parameters and Hamiltonian specified in the quantum system, and the construction of the parameterized quantum circuit in QHL includes decomposing the unitary operator $U(\theta)$. 
	Loss function $Loss(\theta)$ of QHL in Eq.(\ref{Eq:QHL loss}) is used to estimate the difference between the intermediate quantum state $|\psi_{t}(\theta)\rangle$ and the final quantum state with the target Hamiltonian $H_{target}$,
	\begin{equation}
		\label{Eq:QHL loss}
		Loss(\theta) = \langle\psi_{t}(\theta)|H_{target}|\psi_{t}(\theta)\rangle .
	\end{equation}
	By minimizing $Loss(\theta)$ and updating the parameters in parameterized quantum circuits, QHL can precisely represent and learn the Hamiltonian $H_{target}$.

	\section{Quantum Graph Hamiltonian Learning}
	
	\label{section_iii}
	Quantum graph Hamiltonian learning method and its parameterized quantum circuit structure are introduced in this section. 
	At first, the framework of QGHL is presented.
	Then the Hamiltonian and parameterized quantum circuits in QGHL are introduced.
	
	\subsection{QGHL Method}
	
	\begin{figure*}[t]
		\centering
		\includegraphics[width=0.85\textwidth]{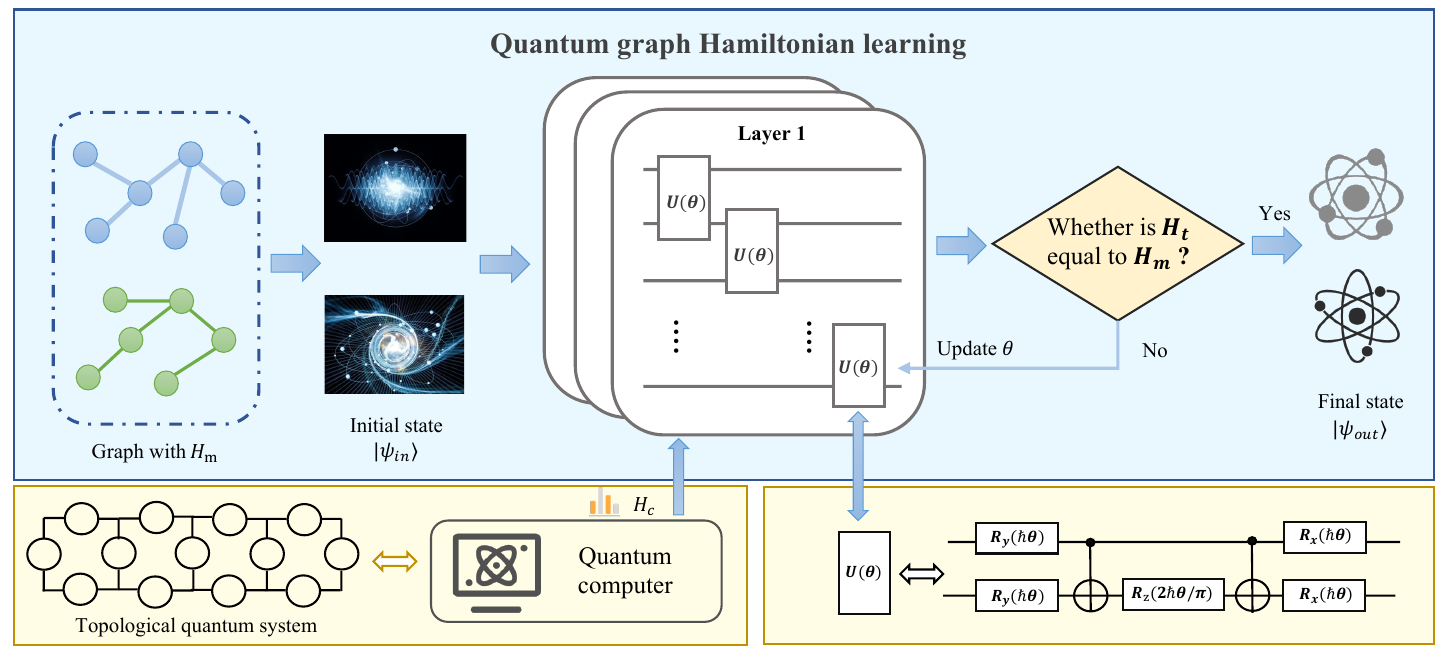}
		\caption{Framework of quantum graph Hamiltonian learning. QGHL is introduced as an approach for linking graphs with the Hamiltonian of a topological quantum system, utilizing the lattice properties of quantum systems to construct parameterized quantum circuits, which eventually produce a quantum state that encapsulates graph information.}
		\label{fig:QGHL}
	\end{figure*}

	Fig. (\ref{fig:QGHL}) illustrates the whole structure of the QGHL method. 
	QGHL is described as a method for mapping a graph to the Hamiltonian of a topological quantum system. 
	It uses the lattice features of a topological quantum system to create parameterized quantum circuits and a final quantum system state including graph information.
	QGHL maps the graph $G$ into the quantum state $|\psi_{out}\rangle$ by associating the graph $G$ with the Hamiltonian $H_m$ of the quantum topological system and optimizing the parameters of the quantum circuit.
	In graph theory, a graph $G$ can be defined as a topological unit often denoted as $G = (V, A)$, where $V$ represents the collection of nodes and $A$ represents the edge adjacency matrix that encodes the relationships between different nodes. 
	A topological quantum system in quantum mechanics refers to a quantum system that exhibits periodic boundary conditions. 
	This system can be contained in a square lattice as shown in Fig. (\ref{fig:QGHL}).
	Creating a mapping connection between the graph $G$ and the Hamiltonian $H_m$ of the quantum topological system makes it possible to encode information from graph $G$ into the quantum state $|\psi_{out}\rangle$. 
	The quantum state in QGHL experiences a system evolution in the quantum environment that consists of the topological quantum system, the initial quantum state$|\psi_{in}\rangle$, and Hamiltonian $H_{m}$, $H_{c}$.
	
	Fig. (\ref{fig:QGHL}) depicts quantum gates made up of decomposable unitary operators $U(\theta)$ with parameters. 
	The parameterized quantum circuits in QGHL comprise quantum gates with multilayer repeating.
	The quantum gates in the circuit excite the initial quantum state, producing the intermediate state $|\psi_{t}(\theta)\rangle$ and the intermediate Hamiltonian $H_t$. 
	The gap of the intermediate Hamiltonian $H_t$ and the Hamiltonian $H_m$ of the quantum topological system is compared to determine if $|\psi_{t}(\theta)\rangle$ is the predicted quantum state $|\psi_{out}\rangle$ that correctly encodes the graph $G$.
	If the difference $Gap=|H_t - H_m|$ exceeds a minimal value $\delta$, the parameters of the quantum circuit depicted in Fig. (\ref{fig:QGHL}) must be optimized. 
	The optimization employs the gradient descent approach to train and update the parameters in quantum circuits by constructing the loss function $Loss(\theta)=\langle \psi_{t}(\theta) |H_m| \psi_{t}(\theta) \rangle$. 
	Conversely, if the difference $Gap=|H_t - H_m|$ is less than $\delta$, it is concluded that the characteristics of the graph $G$ have been encoded into the quantum state, and the current quantum state $|\psi_{out}\rangle$ is produced as an output result. 
	Therefore, QGHL accomplishes the mapping of graph $G$ into the quantum state $|\psi_{out}\rangle$ by learning the topological Hamiltonian $H_m$.

	\subsection{Hamiltonian of QGHL}
	Hamiltonian is an operator that characterizes the energy level of a quantum system. 
	Pauli operators characterize the spin operations of quantum states on the lattice in topological quantum systems and are pivotal operators in constructing the Hamiltonian. 
	The mathematical representation of Pauli operators comprises three $2 \times 2$ matrices, shown in Eq.(\ref{Eq: Paoli}),
	\begin{equation}
		\label{Eq: Paoli}
		\sigma_x = \begin{pmatrix} 0 & 1 \\ 1 & 0 \end{pmatrix}, \quad \sigma_y = \begin{pmatrix} 0 & -i \\ i & 0 \end{pmatrix}, \quad \sigma_z = \begin{pmatrix} 1 & 0 \\ 0 & -1 \end{pmatrix},
	\end{equation}
	where $\sigma_x$, $\sigma_y$, $\sigma_z$ represent the Pauli-X, Pauli-Y, and the Pauli-Z operators respectively. 
	These Pauli operators represent the Pauli-X, Pauli-Y, and Pauli-Z quantum gates in quantum circuits respectively. 
	In a topological quantum system that includes Pauli operations, the Hamiltonian is often expressed as a combination of these Pauli operators and coupling constants that are used to describe the interactions between qubits.
	
	Defining the correspondence between graph $G$ and the Hamiltonian $H_m$ in a topological quantum system is essential for representing and learning graph $G$ via quantum circuits.
	The lattice structure features of the topological quantum system in Fig. (\ref{fig:QGHL}) allow for the establishment of an association between the Hamiltonian $H_m$ in the topological quantum system and graph $G$.
	Spin states of various quantum states in the topological quantum system can be used to identify the nodes $V$ of graph $G$.
	The relationship between the lattice of the topological quantum system and nodes of the graph $G$ has been built,
	The edge and adjacency matrix $A_{ij}$ of graph $G$ can be found by analyzing the interactions between these quantum states.
	Eq.(\ref{Eq: Hamiltonian-mapping}) describes the mapping Hamiltonian $H_m$ of the topological quantum system:
	\begin{equation}
		\label{Eq: Hamiltonian-mapping}
		H_m = \sum_{i,j}^V A_{ij} (J_x \sigma_x^i \sigma_x^j + J_y \sigma_y^i \sigma_y^j + J_z \sigma_z^i \sigma_z^j),
	\end{equation}
	where $A_{ij}$ is the adjacency matrix of graph $G$, and $J_x, J_y, J_z$ are constant coefficients representing the coupling constants of Pauli operators at different positions. 
	QGHL optimizes and updates the parameters of the quantum circuit in Fig. (\ref{fig:QGHL}) by minimizing the loss function that is composed of the Hamiltonian $H_m$ and the intermediate quantum state $|\psi_{t}(\theta)\rangle$. 
	The gap between the Hamiltonian $H_t$ generated during the system evolution and the mapping Hamiltonian $H_m$ determines whether the quantum system evolved through the quantum circuit is the final state.

	\subsection{Quantum circuit of QGHL}

	The quantum gates in the parameterized quantum circuit of QGHL can be acquired by the decomposition of the unitary operator $U(\theta)$, which contains parameters. 
	The unitary operator $U(\theta)$ provides the mathematical expression of quantum gates in the field of quantum computing, shown in Eq.(\ref{Eq:QGHL-Unitary operators}).
	\begin{equation}
		\label{Eq:QGHL-Unitary operators}
		U(\theta) = e^{-\frac{iH_{c}\theta}{\hbar}},
	\end{equation}
	where $\hbar$ represents the Planck constant, and $H_c$ denotes the Hamiltonian that makes up the quantum circuit in QGHL. 
	The Hamiltonian $H_c$ presents the behavior of the quantum system, and the unitary operator $U(\theta)$ represents the changes in the quantum state over time. 
	Quantum gates are used in quantum computing to manipulate qubits and perform various operations, which are a special implementation of the unitary operator $U(\theta)$.
	By designing and using the Hamiltonian, it is possible to build specialized quantum gates that can modify quantum states and do specified computing tasks.
	Eq.(\ref{Eq: QGHL-HC}) depicts the Hamiltonian that composes the unitary operator $U(\theta)$ in QGHL quantum circuits.
	\begin{equation}
		\label{Eq: QGHL-HC}
		H_c = \sum_{n=1}^N \sigma_y^n + \sum_{n=1}^N \left( \frac{1}{2} I - \frac{\pi}{4} \sigma_z^n \otimes \sigma_z^{n+1} \right) + \sum_{n=1}^N \sigma_x^n,
	\end{equation}
	where $\sigma_x$ is the Pauli-X operator, $\sigma_y$ is the Pauli-Y operator, $\sigma_z$ is the Pauli-Z operator, $I$ is the identity matrix, $\otimes$ is the tensor product operation, and $n=1, 2, \ldots, N$ is the number of qubits.
	By substituting Eq.(\ref{Eq: QGHL-HC}) into Eq.(\ref{Eq:QGHL-Unitary operators}), the unitary operator $U(\theta)$ of QGHL is obtained as shown in Eq.(\ref{Eq: QGHL-U-01}).
    \begin{equation}
		\label{Eq: QGHL-U-01}
		U\!(\theta)\!\!=\!exp(-\frac{i \theta}{\hbar}\!\!\left[\sum_{n=1}^N \sigma_y^n\!\!+\!\!\sum_{n=1}^N \left(\frac{1}{2}I\!-\!\frac{\pi}{4} \sigma_z^n \otimes \sigma_z^{n+1}\right)\!\!+\!\!\sum_{n=1}^N \sigma_x^n\!\right]\!).
	\end{equation}
	Eq.(\ref{Eq: QGHL-U-02}) is obtained by decomposing $U(\theta)$ in Eq.(\ref{Eq: QGHL-U-01}). 
	\begin{equation}
		\label{Eq: QGHL-U-02}
		\begin{gathered}
			U\!(\theta)\!\!=\!exp(-\frac{i \theta}{\hbar}\!\!\left[\sum_{n=1}^N \sigma_y^n\!\!+\!\!\sum_{n=1}^N \left(\frac{1}{2}I\!-\!\frac{\pi}{4} \sigma_z^n \otimes \sigma_z^{n+1}\right)\!\!+\!\!\sum_{n=1}^N \sigma_x^n\!\right]\!)\\
			=exp(-\frac{i \theta}{\hbar} \sum_{n=1}^N \sigma_n^y) \times exp(-\frac{i \theta}{\hbar} \sum_{n=1}^N \frac{I}{2}) \\
			\times exp(\frac{i \pi \theta}{4 \hbar} \left[\sum_{n=1}^N \sigma^n_z \otimes \sigma^{n+1}_z\right]) \times exp(-\frac{i \theta}{\hbar} \sum_{n=1}^N \sigma^n_x) \\
			=\prod_{n=1}^N e^{-\frac{i\sigma^n_y \theta}{\hbar} } \times \prod_{n=1}^N e^{-\frac{iI \theta}{2 \hbar} } \times \prod_{n=1}^N e^{\frac{i \pi \sigma^n_z \otimes \sigma^{n+1}_z \theta}{4 \hbar}} \times \prod_{n=1}^N e^{-\frac{i\sigma^n_x \theta}{\hbar} }.
		\end{gathered}
	\end{equation}
	Then, Eq.(\ref{Eq: QGHL-U-03}) is obtained by decomposing $ e^{\frac{i \pi \sigma^n_z \otimes \sigma^{n+1}_z \theta}{4 \hbar}}$.
	{
		\begin{equation}
			\begin{gathered}
				\label{Eq: QGHL-U-03}
				\prod_{n=1}^N e^{\frac{i \pi \sigma^n_z \otimes \sigma^{n+1}_z \theta}{4 \hbar}}=\prod_{n=1}^N\left[\begin{array}{cccc}
					e^{\frac{i \pi \theta}{4 \hbar}} & 0 & 0 & 0 \\
					0 & e^{\frac{-i \pi \theta}{4 \hbar}} & 0 & 0 \\
					0 & 0 & e^{\frac{-i \pi \theta}{4 \hbar}} & 0 \\
					0 & 0 & 0 & e^{\frac{i \pi\theta}{4 \hbar}}
				\end{array}\right] \\
				=\prod_{n=1}^N\left[\begin{array}{llll}
					\! \!1 &\! \! 0 &\! \! 0 &\! \! 0 \\
					\! \!0 &\! \! 1 &\! \! 0 &\! \! 0 \\
					\! \!0 &\! \! 0 &\! \! 0 & \! \!1 \\
					\! \!0 & \! \!0 &\! \! 1 &\! \! 0
				\end{array}\right]\! \!\left[\! \!\begin{array}{cccc}
					\! \!e^{\frac{i \pi \theta}{4 \hbar}} &\! \!\! \! 0 & \! \!\! \!0 &\! \!\! \! 0 \\
					\! \!0 &\! \!\! \! e^{\frac{-i \pi \theta}{4 \hbar}} & \! \!\! \!0 &\! \!\! \! 0 \\
					\! \!0 &\! \!\! \! 0 &\! \!\! \! e^{\frac{-i \pi \theta}{4 \hbar}} &\! \! \! \!0 \\
					\! \!0 &\! \!\! \! 0 &\! \!\! \! 0 & \! \!\! \!e^{\frac{i \pi \theta}{4 \hbar}}
				\end{array}\right]\! \!\left[\! \!\begin{array}{llll}
					\! \!1 & \! \!0 &\! \! 0 &\! \! 0 \\
					\! \!0 & \! \!1 & \! \!0 & \! \!0 \\
					\! \!0 &\! \! 0 & \! \!0 &\! \! 1 \\
					\! \!0 &\! \! 0 &\! \! 1 &\! \! 0
				\end{array}\right] \\
				=\prod_{n=1}^N \text { CNOT Gate } \cdot R_z\left(\frac{2 \hbar}{\pi}\theta \right) \cdot \text { CNOT Gate }.
			\end{gathered}
		\end{equation}
	}
	
\begin{figure*}[bt]
		\[ \Qcircuit @C=1em @R=.7em {
			&\qw& \qw& \qw& \qw& \gate{R_{y}(\hbar\theta)} &   \ctrl{1}& \qw & \ctrl{1} & \gate{R_{x}(\hbar\theta)}& \qw& \qw& \qw&  \qw& \qw& \qw& \qw&\qw&\qw& \qw& \qw& \qw& \meter& \qw\\
			&\qw& \qw& \qw& \qw& \gate{R_{y}(\hbar\theta)} &  \targ& \gate{R_{z}(\frac{2\hbar\theta}{\pi})} & \targ & \gate{R_{x}(\hbar\theta)}&\qw& \qw& \qw& \qw& \gate{R_{y}(\hbar\theta)}&   \ctrl{1}& \qw & \ctrl{1} & \gate{R_{x}(\hbar\theta)}& \qw& \qw& \qw& \meter& \qw\\
			&\qw& \qw& \qw& \qw& \qw &  \qw& \qw & \qw & \qw&\qw& \qw& \qw& \qw& \gate{R_{y}(\hbar\theta)}& \targ& \gate{R_{z}(\frac{2\hbar\theta}{\pi})} & \targ & \gate{R_{x}(\hbar\theta)}& \qw&\qw& \qw& \meter& \qw\\
		} \]
		\caption{The example of the quantum circuits in QGHL. Quantum circuits of QGHL are made up of a 3-qubit quantum system with $1$ repetition layers.}
		\label{fig:QGHL circuit}
	\end{figure*}
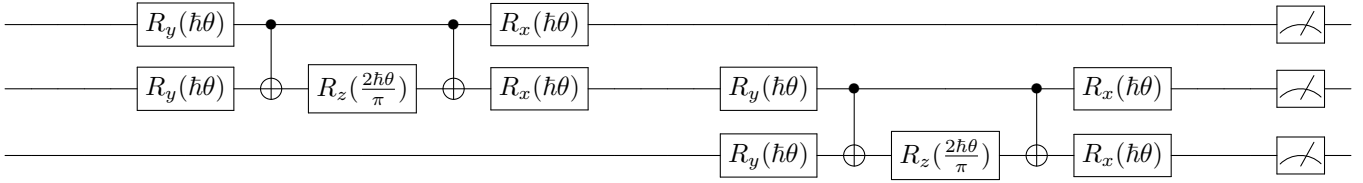

	Finally, the quantum gates that comprise the QGHL quantum circuit are presented in Eq. (\ref{Eq:QGHL-Gate}),
	\begin{equation}
		\label{Eq:QGHL-Gate}
		\begin{split}
			U(\theta)= R_y( \hbar \theta) \! \!\cdot \! \!\text { CNOT }\! \! \cdot\! \! R_z\left(\frac{2 \hbar}{\pi} \theta\right)
			\! \!\cdot\! \! \text { CNOT }\! \! \cdot\! \! R_x( \hbar \theta).
		\end{split}
	\end{equation}
	The QGHL quantum circuit is made up of three quantum rotation gates: $R_y$, $R_z$,$R_x$ and $CNOT$ gate. 
	Fig. (\ref{fig:QGHL circuit}) depicts the QGHL quantum circuit with 3 qubits and $1$ repetition layers.
	
	\section{Quantum Graph Hamiltonian Neural Network}
	
	\label{section_iiii}
	
	In this section, the structure of QGHNN is presented, which depends on QGHL.
	QGHNN updates the parameters in quantum neurons by minimizing the loss function $Loss(\theta)$, which is composed of the Hamiltonian $H_{m}$ derived from graph $G$ and the vector product of intermediate quantum states $|\psi_t(\theta)\rangle$. 
	QGHNN enables the quantum neural network to learn information for the graph.
	First, we give the framework of QGHNN. 
	Then, we introduce an algorithm that is based on QGHNN.
	
	\subsection{Framework of QGHNN}

\begin{figure}[t]
		\centering
		\includegraphics[width=0.48\textwidth]{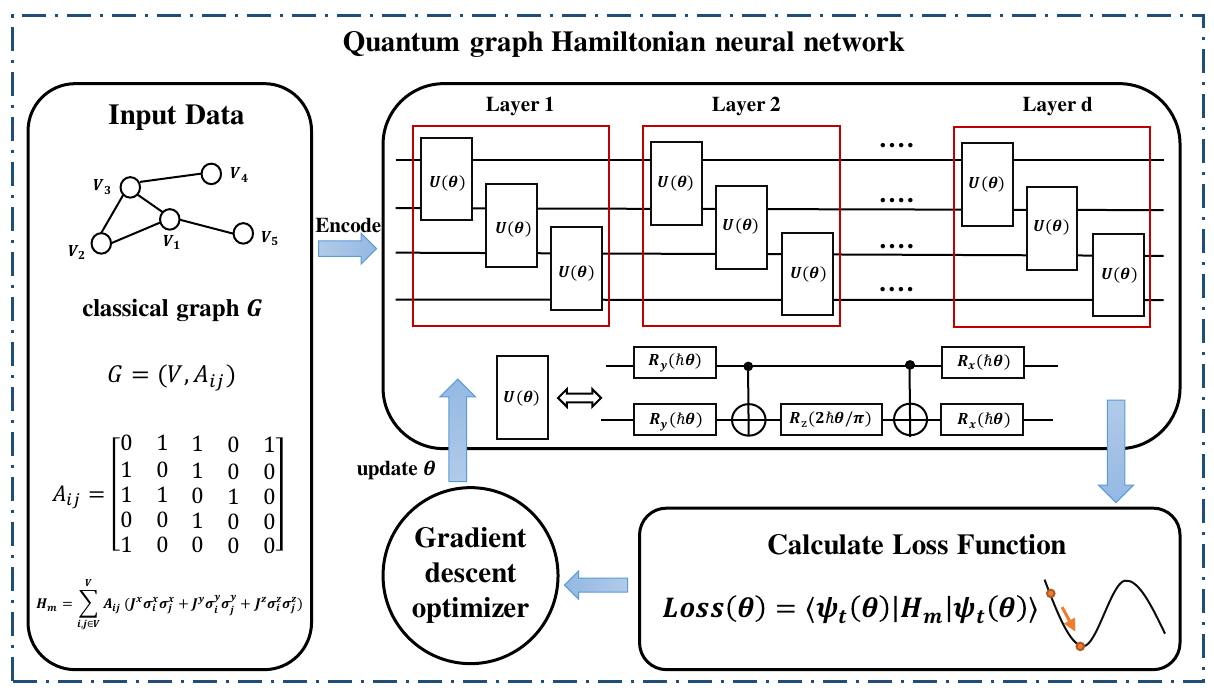}
		\caption{Quantum graph Hamiltonian neural network. The QGHNN model consists of four components: input data, neuron units of the quantum neural network, loss function computation, and parameter updating mechanism based on gradient descent.}
		\label{fig:QGHNN}
	\end{figure}

	Fig. ({\ref{fig:QGHNN}}) illustrates the structural framework of QGHNN. 
	The QGHNN model consists of four components: input data, neuron units of the quantum neural network, loss function computation, and parameter updating mechanism based on gradient descent.
	The input data consists of the classical graph $G$ and the mapping Hamiltonian $H_m$ derived from graph $G$. 
	Graph $G$ is defined as $G = (V, A_{ij})$, where $V$ represents the node information and $A_{ij}$ is the adjacency matrix of graph $G$.
	The components of $H_m$ as shown in Eq.(\ref{Eq: Hamiltonian-mapping}) include the coupling constants $J_x, J_y, J_z$, the adjacency matrix $H_m$, and the Pauli operators $\sigma_x, \sigma_y, \sigma_z$.
	The QGHNN uses parameterized quantum circuits that are constructed by QGHL. 
	Each neuron consists of unitary operator $U(\theta)$, which are built of quantum gates like $R_y( \hbar \theta)$, $R_x( \hbar \theta)$, $R_z(\frac{2\hbar\theta}{\pi})$, and CNOT gates.
	Every individual neuron enhances the interaction between neighbouring qubits.
	The loss function $Loss(\theta)$ consists of the mapping Hamiltonian $H_m$ designed based on graph $G$ and the vector product of intermediate quantum states $|\psi_{t}(\theta)\rangle$, as shown in Eq.(\ref{Eq:QGHNN-loss}),
	\begin{equation}
		\label{Eq:QGHNN-loss}
		Loss(\theta) = \langle \psi_t(\theta) | H_m | \psi_t(\theta) \rangle.
	\end{equation}
	The neurons in QGHNN are updated by minimizing the loss function $Loss(\theta)$.
	The parameters $\theta$ of quantum circuits are updated using the gradient descent approach by the QGHNN.  
	The update procedure is described by the Eq.(\ref{Eq:QGHNN-gradient}).
	\begin{equation}
		\label{Eq:QGHNN-gradient}
		\theta^* \leftarrow \theta - R \frac{\partial Loss(\theta)}{\partial \theta},
	\end{equation}
	where $\theta^*$ is the updated parameter and $R$ is the learning rate \cite{ref.zhang.yang2023sagn}.

	\subsection{Algorithm of QGHNN}
	
	Algorithm \ref{algorithm} presents the pseudo-code of QGHNN. 
	The algorithm takes inputs such as $V, A_{ij}, H_m, H_c, R, n, d$.
	The QGHNN algorithm depends on the values of the following parameters: the node count $V$ of graph $G$, the adjacency matrix $A_{ij}$ of graph $G$, the mapping Hamiltonian $H_m$ of graph $G$, the Hamiltonian $H_c$ of the quantum circuits that decompose the unitary operators in QGHNN, the learning rate $R$, qubits number $n$, and layers $d$.
	The output comprises many assessment metrics, including the loss function, mean squared error (MSE), cosine similarity, Frobenius norm, and correlation coefficient. 
	These metrics are used to measure the differences between the calculated results and the input graph $G$ information.
	The MSE is a metric that quantifies the average squared difference between two data sets, which elucidates the divergence between the predicted values and the actual values of an algorithm.
	A minor MSE indicates a higher similarity between the predicted outcomes and the actual values \cite{ref.tetic.LXL.liu2024reinforcement}.
	Cosine similarity quantifies the degree of similarity in direction between two matrices or vectors, where the magnitude of the cosine similarity directly corresponds to the degree of directional similarity exhibited by the two matrices or vectors \cite{ref.LXL.zhang2024decouple}.
	The Frobenius norm quantifies the magnitude of the discrepancy between two matrices and the result of a correlation coefficient that is closer to 1 indicates a better positive correlation between the predicted outcomes and actual values, suggesting that both variables tend to change in the same direction \cite{ref.zhang.chen2024qgrl}.
	Algorithm \ref{algorithm} initializes the parameters $\theta$ in the quantum circuits of QGHNN randomly. 
	The definitions of Hamiltonians $H_m$ and $H_c$ may be found in Eq.(\ref{Eq: Hamiltonian-mapping}) and Eq.(\ref{Eq: QGHL-HC}), respectively.
	The parameters $\theta$ in the quantum circuits of QGHNN are updated by the computation of the partial derivatives of the loss function $Loss(\theta)$ and subsequent minimization.
    
	\begin{algorithm}[H]
		\caption{The Algorithm of QGHNN.}
		\label{algorithm}
		\begin{algorithmic}
			
			\REQUIRE  $V, A_{ij}, H_m, H_c, R, n, d$;
			\ENSURE Loss function, MSE, cosine similarity, Frobenius norm, correlation coefficient.
			\STATE \textit{Step} 1: Input classical graphs data
			\STATE \textit{Step} 2: Prepare the initial quantum system with random parameter $\theta$.
			\STATE \textit{Step} 3: Generate the quantum circuits of QGHNN and its neuron units.
			\STATE \textit{Step} 4: Calculate the loss function $Loss(\theta)$ and update the parameter $\theta$.
			\STATE For { $j<\xi$ :}
			\STATE
			\STATE \hspace{0.5cm}$\frac{\partial Loss(\theta_j)}{\partial \theta_j} = \frac{\partial Loss(\theta_j + \Delta_j) - \partial Loss(\theta_j - \Delta_j)}{2 \Delta_j}$;
			\STATE
			\STATE \hspace{0.5cm}$\theta_{j+1}^* \leftarrow \theta_j - R \frac{\partial Loss(\theta_j)} {\partial \theta_j}$.
			
		\end{algorithmic}
	\end{algorithm}
	
	\section{Experiment and Discussion}
	
	QGHNN has the potential to represent and learn graph $G$ on NISQ devices and can be integrated with deep learning and other language models for broader applications in the future. 
	We aim to ensure that QGHNN can perform Hamiltonian $H_{m}$ of topological quantum systems.
	After employing several experiments, we discovered that QGHNN can effectively acquire information from classical graph $G$ on NISQ devices. 
	This section presents experiments and discussions on QGHNN and QGHL. 
	First, we provide experiments in which quantum computers are used to represent and learn classical graph $G$, demonstrating the advantages of QGHNN in learning graph data over other quantum machine learning models. 
	Subsequently, we conduct a performance analysis of QGHNN, QGHL, and other QML models to illustrate the superiority of the QGHNN.

	\subsection{Experiment setting}
	
	\begin{figure}[t]
		\centering
		\includegraphics[width=0.48\textwidth]{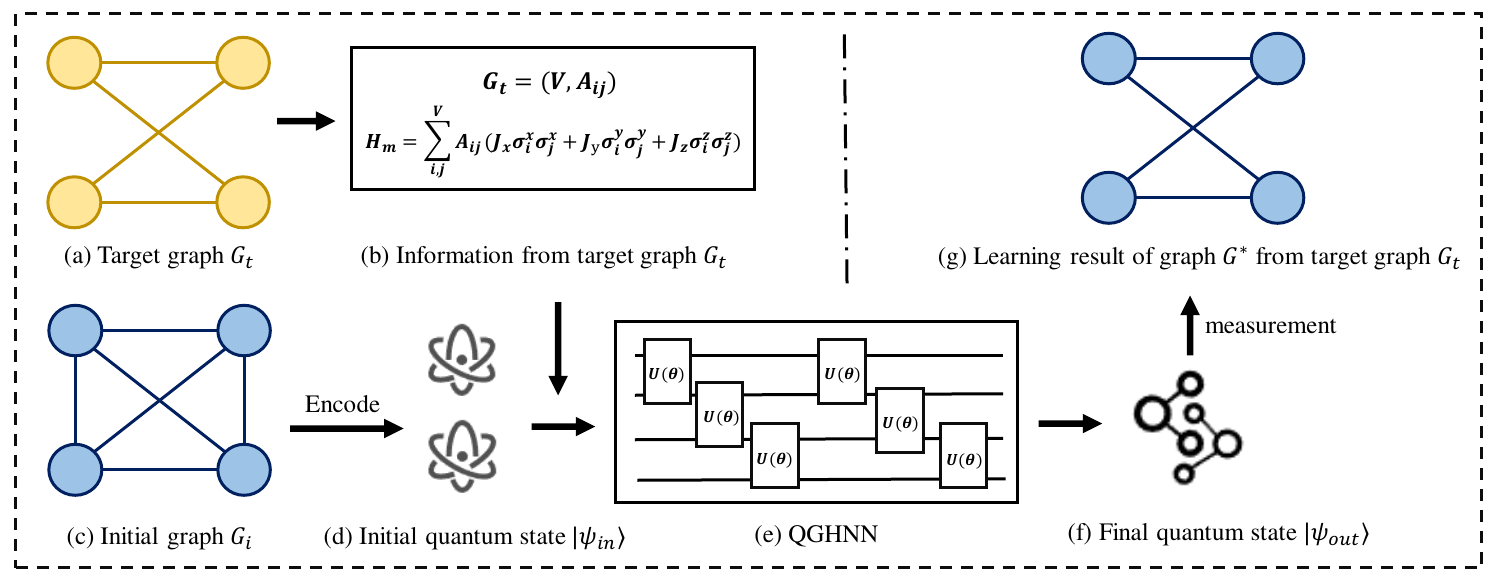}
		\caption{Experiment setting of QGHNN. QGHNN learns the target graph $G_t$ by encoding the initial fully connected graph $G_i$ and training the parameters.}
		\label{fig:Experiment setting of QGHNN}
	\end{figure}
	
	Fig. (\ref{fig:Experiment setting of QGHNN}) depicts the quantum experiment setting for learning graph data by QGHNN and the target graph $G_t$ has four nodes.
	Fig. (\ref{fig:Experiment setting of QGHNN})(a) depicts the target graph $G_t$ that should be learned and the direct extraction of important information in $G_t$ are $G_t = (V, A_{ij})$ and the mapped Hamiltonian $H_m$. 
	Fig. (\ref{fig:Experiment setting of QGHNN})(b) shows the data obtained from the target graph $G_t$, which is used as inputs of the QGHNN model.
    QGHNN uses the quantum amplitude coding to encode the graph.
    The quantum state needs to satisfy the normalization condition, so QGHNN needs to normalize the elements of the adjacency matrix $ A_{ij} $ as shown in Eq. (\ref{Eq.a01}).
   \begin{equation}
   \label{Eq.a01}
       A'_{ij} = \frac{A_{ij}}{\sqrt{\sum_{i,j} A_{ij}^2}}.
   \end{equation}
  We map the normalized matrix elements $ A'_{ij} $ to the amplitude of the quantum state.
  For a $ n \times n $ graph $G_t = (V, A_{ij})$, the input quantum state of QGHNN is shown in Eq. (\ref{Eq.input01}).
   \begin{equation}
   \label{Eq.input01}
       \lvert \psi \rangle = \sum_{i,j} A'_{ij} \lvert i \rangle \lvert j \rangle.
   \end{equation}
	Fig. (\ref{fig:Experiment setting of QGHNN})(c) represents a fully connected graph with the nodes as the target graph $G_t$. 
	Encoding the initial graph $G_i$ in Fig. (\ref{fig:Experiment setting of QGHNN})(c) into the initial quantum state $|\psi_0\rangle$ creates a quantum environment of QGHNN. 
	The quantum state $|\psi_0\rangle$ represents the starting graph $G_i$, which develops through the QGHNN model to produce the final quantum state $|\psi_{out}\rangle$. 
	Fig. (\ref{fig:Experiment setting of QGHNN})(g) depicts the learning result graph $G^*$, which is generated by the final quantum state.
	
	The experiment uses MSE, cosine similarity, Frobenius norm, and correlation coefficient to assess the similarity between Fig. (\ref{fig:Experiment setting of QGHNN})(a) and Fig. (\ref{fig:Experiment setting of QGHNN})(g).
	Among these, MSE compares the difference between Fig. (\ref{fig:Experiment setting of QGHNN})(g) and the actual graph $G_t$. 
	Cosine similarity is used to assess the direction similarity of the adjacency matrix vectors, and the Frobenius norm analyzes the total difference of graphs.
	The correlation coefficient represents the positive correlation between Fig. (\ref{fig:Experiment setting of QGHNN})(g) and Fig. (\ref{fig:Experiment setting of QGHNN})(a). 
	To illustrate the advantage of QGHNN in learning graph data, we compared it to several typical quantum machine learning algorithms, including VQE\cite{ref.VQE.du2022quantum}, QAOA\cite{ref.QAOA.guerreschi2019qaoa}, and QNN\cite{ref.zhao2024gqhan}.
	We use the PennyLane quantum experimental platform to do the experiments.
	Three different experiments are designed to prove the advantages of QGHNN in learning graphs:
	
	\begin{enumerate}
		\item \textbf{Experiment (01)} : 4-qubit QGHNN to learn target graph $G_{t_1}$ , $G_{t_1} = (4, A_{t_1})$
		\item \textbf{Experiment (02)} : 5-qubit QGHNN to learn target graph $G_{t_2}$ , $G_{t_2} = (5, A_{t_2})$
		\item \textbf{Experiment (03)} : 6-qubit QGHNN to learn target graph $G_{t_3}$ , $G_{t_3} = (6, A_{t_3})$
	\end{enumerate} 
	where the adjacency matrices are shown in Eq.(\ref{Eq.adjacency matrices}).
	{\footnotesize
		\begin{equation}
			\label{Eq.adjacency matrices}
			A_{t_1}=\left[\begin{array}{llll}
				\! \!\! \!0 &\! \!\! \! 1 &\! \!\! \! 0 &\! \!\! \! 1\! \!\! \! \\
				\! \!\! \!1 &\! \!\! \! 0 &\! \!\! \! 1 &\! \!\! \! 0 \! \!\! \!\\
				\! \!\! \!0 &\! \!\! \! 1 &\! \!\! \! 0 &\! \!\! \! 1 \! \!\! \!\\
				\! \!\! \!1 &\! \!\! \! 0 &\! \!\! \! 1 & \! \!\! \! 0\! \!\! \!
			\end{array}\right], A_{t_2}=\left[\begin{array}{lllll}
				\! \!\! \! 0 & \! \!\! \!1 &\! \!\! \! 0 & \! \!\! \!0 &\! \!\! \! 1\! \!\! \! \\
				\! \!\! \!1 &\! \!\! \! 0 & \! \!\! \!1 &\! \!\! \! 0 &\! \!\! \! 0 \! \!\! \!\\
				\! \!\! \!0 &\! \!\! \! 1 & \! \!\! \!0 &\! \!\! \! 1 &\! \!\! \! 0 \! \!\! \!\\
				\! \!\! \!0 &\! \!\! \! 0 &\! \!\! \! 1 & \! \!\! \!0 & \! \!\! \!1 \! \!\! \!\\
				\! \!\! \! 1 & \! \!\! \!0 & \! \!\! \!0 & \! \!\! \!1 &\! \!\! \! 0\! \!\! \!
			\end{array}\right]
			A_{\mathrm{t}_3}=\left[\begin{array}{llllll}
				\! \!\! \!0 &\! \!\! \! 0 &\! \!\! \! 0 &\! \!\! \! 1 &\! \!\! \! 1 &\! \!\! \! 0\! \!\! \! \\
				\! \!\! \!0 &\! \!\! \! 0 & \! \!\! \!0 &\! \!\! \! 1 &\! \!\! \! 0 & \! \!\! \!1 \! \!\! \!\\
				\! \!\! \! 0 &\! \!\! \! 0 &\! \!\! \! 0 & \! \!\! \!0 & \! \!\! \!1 &\! \!\! \! 1\! \!\! \! \\
				\! \!\! \!1 &\! \!\! \! 1 & \! \!\! \!0 &\! \!\! \! 0 &\! \!\! \! 0 &\! \!\! \! 0\! \!\! \! \\
				\! \!\! \! 1 & \! \!\! \!0 & \! \!\! \!1 & \! \!\! \!0 & \! \!\! \!0 &\! \!\! \! 0 \! \!\! \!\\
				\! \!\! \! 0 & \! \!\! \!1 &\! \!\! \! 1 &\! \!\! \! 0 &\! \!\! \! 0 & \! \!\! \!0\! \!\! \!
			\end{array}\right].
		\end{equation}
	}
	
	In Experiments (01), (02), and (03), we create comparative experiments with various quantum machine learning models. 
	Experiments (01), (02), and (03) show that the complexity of the target graph $G_t$ rises with the number of qubits, demonstrating that QGHNN can apply quantum computing to achieve superior learning capabilities. 
	The experimental findings demonstrate that QGHNN can accurately learn graph representations.

	\subsection{Experimental Results and Discussion}
    \begin{figure}[t]
		\centering
		\includegraphics[width=0.48\textwidth]{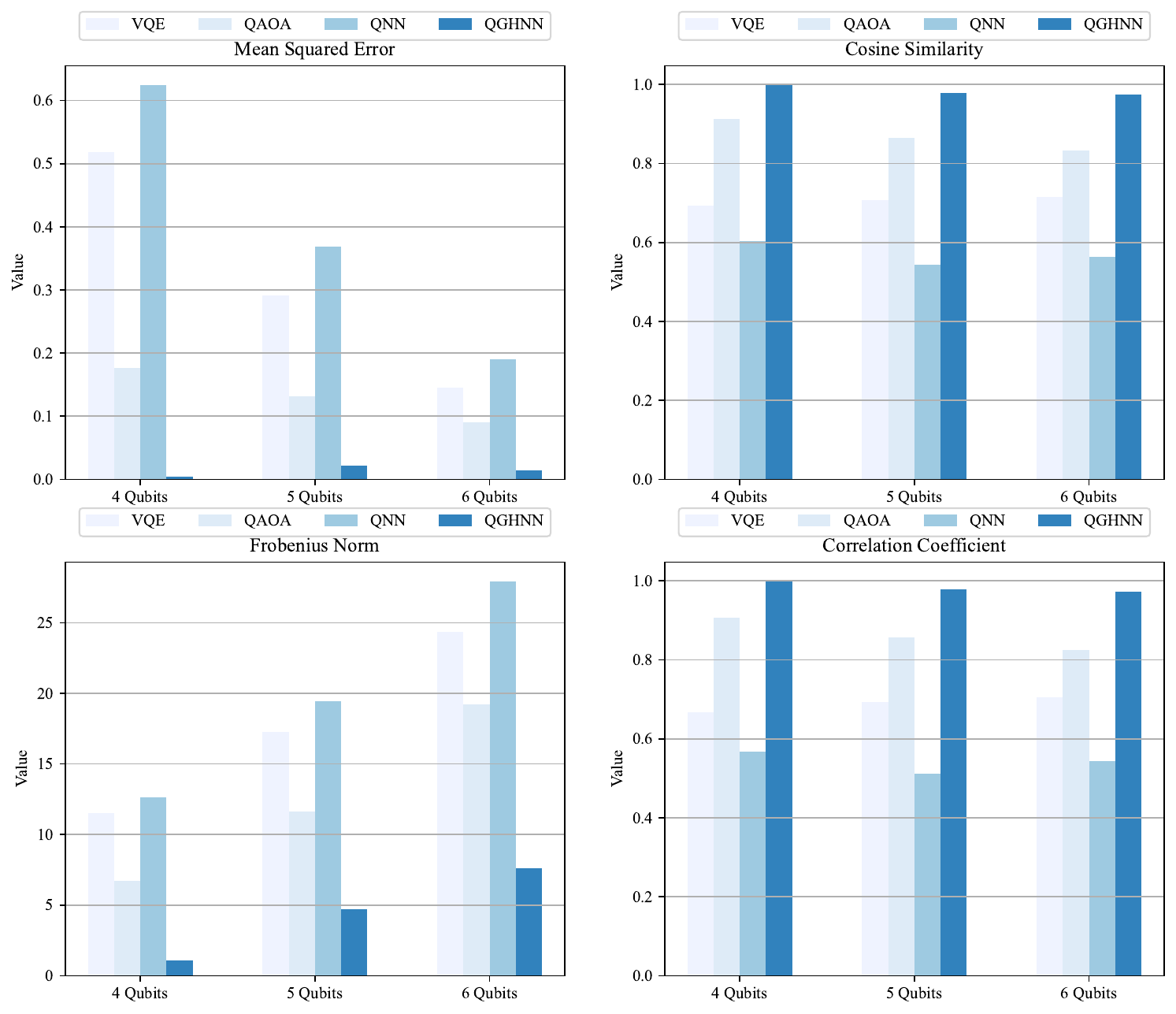}
		\caption{Experimental comparison results of QGHNN and other different quantum machine learning models on 4 qubits, 5 qubits, and 6 qubits respectively.}
		\label{fig:comparation01}
	\end{figure}
	
	\begin{table*}[ht]
		\setlength{\abovecaptionskip}{0cm}  
		\setlength{\belowcaptionskip}{-0.2cm} 
		\centering
		\renewcommand\arraystretch{1.8}
		\caption{Experimental results of different quantum machine learning models in four qubits}
		\setlength{\tabcolsep}{3.0mm}{
			\begin{tabular}{lcccc}
				\toprule
				QML Model & Mean squared error & Cosine Similarity & Frobenius Norm & Correlation Coefficient \\
				\midrule
				VQE \cite{ref.VQE.du2022quantum}   & 0.518 & 0.693 & 11.516 & 0.668 \\
				QAOA \cite{ref.QAOA.guerreschi2019qaoa} & 0.176 & 0.912 & 6.710  & 0.906 \\
				QNN  \cite{ref.zhao2024gqhan} & 0.624 & 0.604 & 12.639 & 0.567 \\
				QGHNN & \textbf{0.004} & \textbf{0.998} & \textbf{1.059}  & \textbf{0.998} \\
				\bottomrule
			\end{tabular}
		}
		\label{tab:Experiment01}
	\end{table*}

	\begin{table*}[ht]
		\setlength{\abovecaptionskip}{0cm}  
		\setlength{\belowcaptionskip}{-0.2cm} 
		\renewcommand\arraystretch{1.8}
		\centering
		\caption{Experimental results of different quantum machine learning models in five qubits}
		\setlength{\tabcolsep}{3.0mm}{
			\begin{tabular}{lcccc}
				\toprule
				QML Model & Mean squared error & Cosine Similarity & Frobenius Norm & Correlation Coefficient \\
				\midrule
				VQE \cite{ref.VQE.du2022quantum}   & 0.291 & 0.708 & 17.273 & 0.693 \\
				QAOA \cite{ref.QAOA.guerreschi2019qaoa}  & 0.131 & 0.865 & 11.604 & 0.857 \\
				QNN  \cite{ref.zhao2024gqhan} & 0.368 & 0.544 & 19.425 & 0.511 \\
				QGHNN & \textbf{0.022} & \textbf{0.979} & \textbf{4.721}  & \textbf{0.978} \\
				\bottomrule
			\end{tabular}
		}
		\label{tab:Experiment02}
	\end{table*}
	
	\begin{table*}[ht]
		\setlength{\abovecaptionskip}{0cm}  
		\setlength{\belowcaptionskip}{-0.2cm} 
		\renewcommand\arraystretch{1.8}
		\centering
		\caption{Experimental results of different quantum machine learning models in six qubits}
		\setlength{\tabcolsep}{3.0mm}{
			\begin{tabular}{lcccc}
				\toprule
				QML Model & Mean squared error & Cosine Similarity & Frobenius Norm & Correlation Coefficient \\
				\midrule
				VQE \cite{ref.VQE.du2022quantum}   & 0.145 & 0.716 & 24.335 & 0.706 \\
				QAOA \cite{ref.QAOA.guerreschi2019qaoa}  & 0.090 & 0.832 & 19.235 & 0.826 \\
				QNN \cite{ref.zhao2024gqhan}  & 0.190 & 0.564 & 27.910 & 0.544 \\
				QGHNN & \textbf{0.014} & \textbf{0.974} & \textbf{7.590}  & \textbf{0.973} \\
				\bottomrule
			\end{tabular}
		}
		\label{tab:Experiment03}
	\end{table*}
	
	\begin{figure*}[ht]
		\centering
		\includegraphics[width=0.9\textwidth]{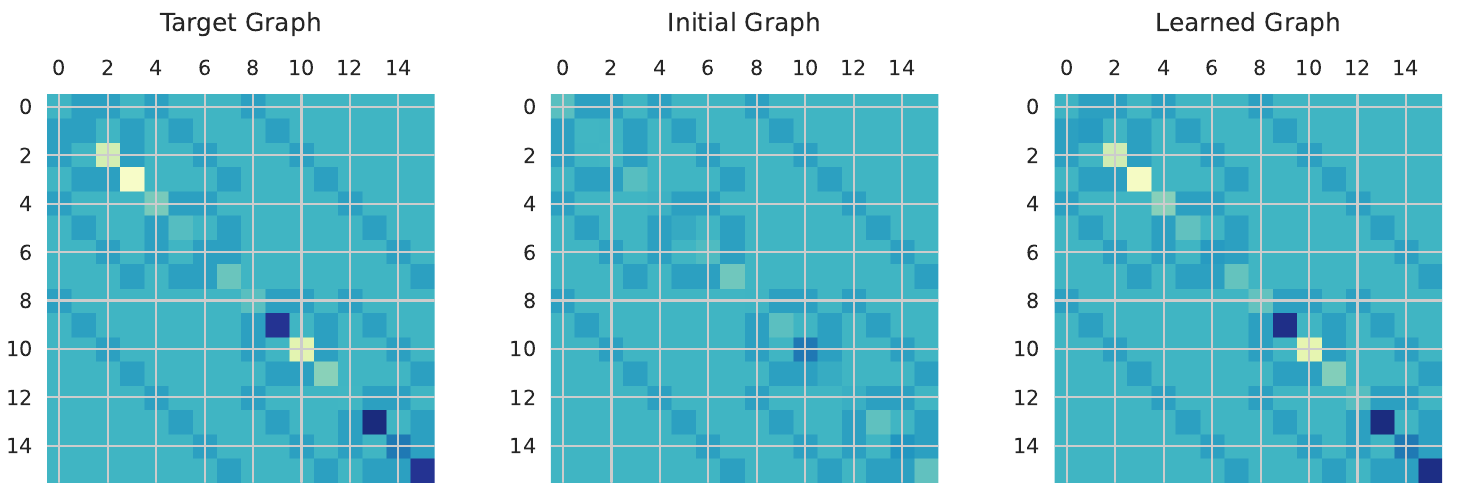}
		\caption{Experiment result of 4-Qubits based on QGHNN model.}
		\label{fig:4-Qubits-QGHNN}
	\end{figure*}
    
	\begin{figure*}[ht]
		\centering
		\includegraphics[width=0.9\textwidth]{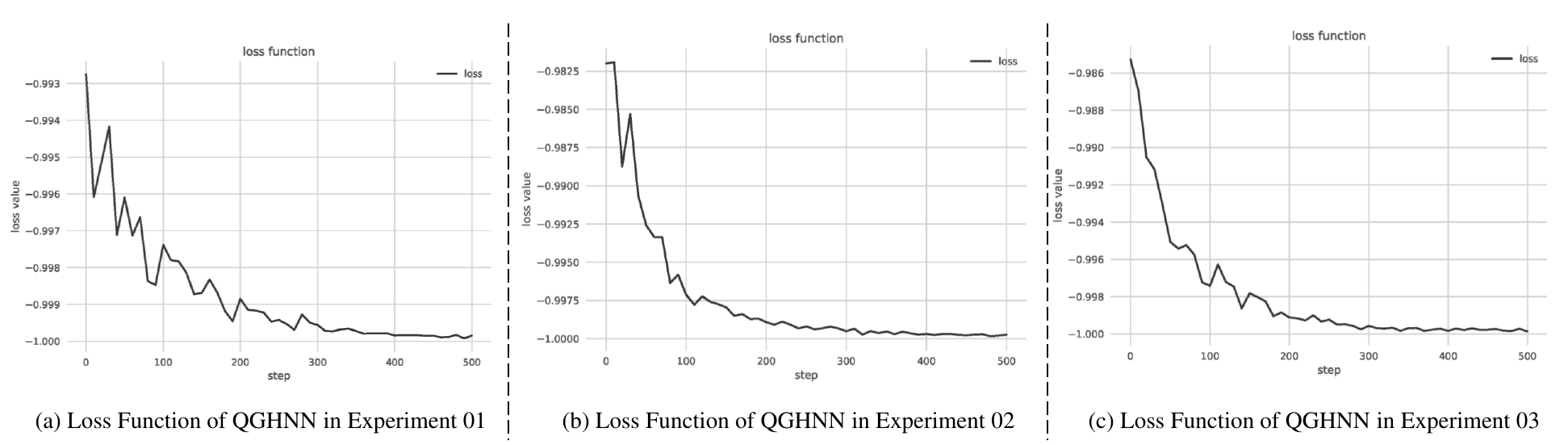}
		\caption{Experiment result of loss function.}
		\label{fig:loss function}
	\end{figure*}

    \begin{figure}[t]
		\centering
		\includegraphics[width=0.48\textwidth]{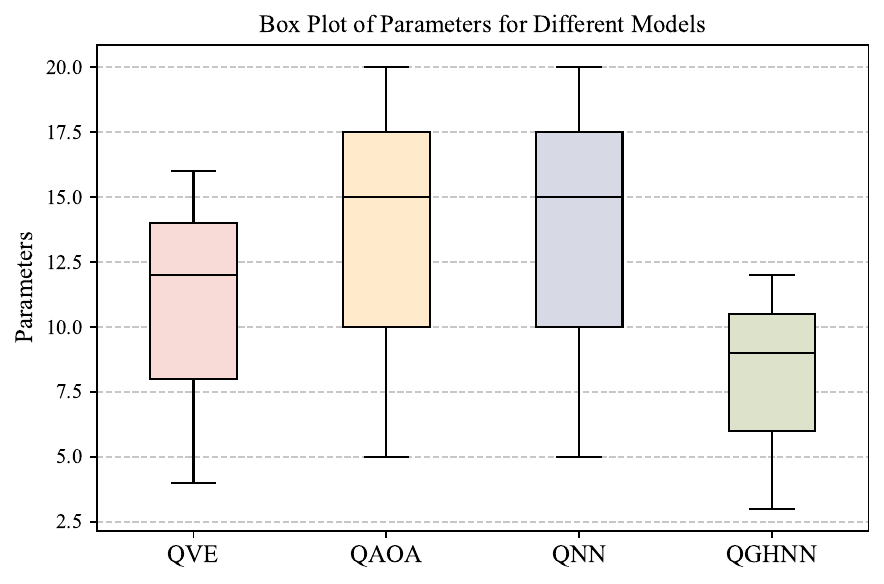}
		\caption{Results about the use of parameters between QGHNN and other different quantum machine learning models.}
		\label{fig:comparation02}
	\end{figure}

    \begin{table}[htpb]
		\renewcommand{\arraystretch}{1.6}
		\caption{Experimental configuration of experiment01.}  
		\label{tab:Experimental configuration01}
		\centering
		\begin{tabular}{p{2.0cm}<{\centering} p{1.2cm}<{\centering} p{1.35cm}<{\centering} p{1.2cm}<{\centering} p{1cm}<{\centering}}  
			\hline  
			Indicators & QVE \cite{ref.VQE.du2022quantum} & QAOA \cite{ref.QAOA.guerreschi2019qaoa} &  QNN \cite{ref.zhao2024gqhan} & QGHNN\\   \hline
			Parameters & 12 &15 &15 & 9\\
                Qubits & 4 & 4 & 4 & 4\\
			Layers &3 &4 &4 & 3  \\
		    Representation Advantages & Weak & Well &Weak & Excellent \\
                Data Encoding & \multicolumn{4}{c}{Quantum Amplitude Encode} \\
			Learning Rate& \multicolumn{4}{c}{0.1} \\
                Batch Size& \multicolumn{4}{c}{10}  \\
                Step& \multicolumn{4}{c}{500} \\
                Optimizer & \multicolumn{4}{c}{Gradient Descent} \\   
                 \hline  
		\end{tabular}  
	\end{table} 

      \begin{table}[htpb]
		\renewcommand{\arraystretch}{1.6}
		\caption{Experimental configuration of experiment02.}  
		\label{tab:Experimental configuration03}
		\centering
		\begin{tabular}{p{2.0cm}<{\centering} p{1.2cm}<{\centering} p{1.35cm}<{\centering} p{1.2cm}<{\centering} p{1cm}<{\centering}}  
			\hline  
			Indicators & QVE \cite{ref.VQE.du2022quantum} & QAOA \cite{ref.QAOA.guerreschi2019qaoa} &  QNN \cite{ref.zhao2024gqhan} & QGHNN\\   \hline
	        Parameters & 16 &20 &20 & 12\\
                Qubits & 5 & 5 & 5 & 5\\
			Layers &6 &6 &5 & 4  \\
		    Representation Advantages & Weak & Well &Weak & Excellent \\
                Data Encoding & \multicolumn{4}{c}{Quantum Amplitude Encode} \\
			Learning Rate& \multicolumn{4}{c}{0.1} \\
                Batch Size& \multicolumn{4}{c}{10}  \\
                Step& \multicolumn{4}{c}{500} \\
                Optimizer & \multicolumn{4}{c}{Gradient Descent} \\   
                 \hline  
		\end{tabular}  
	\end{table} 

      \begin{table}[htpb]
		\renewcommand{\arraystretch}{1.6}
		\caption{Experimental configuration of experiment03.}  
		\label{tab:Experimental configuration02}
		\centering
		\begin{tabular}{p{2.0cm}<{\centering} p{1.2cm}<{\centering} p{1.35cm}<{\centering} p{1.2cm}<{\centering} p{1cm}<{\centering}}  
			\hline  
			Indicators & QVE \cite{ref.VQE.du2022quantum} & QAOA \cite{ref.QAOA.guerreschi2019qaoa} &  QNN \cite{ref.zhao2024gqhan} & QGHNN\\   \hline
			Parameters & 4 &5 &5 & 3\\
                Qubits & 6 & 6 & 6 & 6\\
			Layers &6 &5 &5 & 4  \\
                Representation Advantages & Weak & Well &Weak & Excellent \\
                Data Encoding & \multicolumn{4}{c}{Quantum Amplitude Encode} \\
			Learning Rate& \multicolumn{4}{c}{0.1} \\
                Batch Size& \multicolumn{4}{c}{10}  \\
                Step& \multicolumn{4}{c}{500} \\
                Optimizer & \multicolumn{4}{c}{Gradient Descent} \\   
                 \hline  
		\end{tabular}  
	\end{table} 
	
	Fig. (\ref{fig:comparation01}) introduces the experimental comparison results of QGHNN and other different quantum machine learning models. 
    Table. (\ref{tab:Experiment01}) shows the assessment metrics results from Experiment (01), which was done by employing the PennyLane experimental quantum platform. 
        Table. (\ref{tab:Experimental configuration01}) shows the parameter settings for Experiment (01).
	Fig. (\ref{fig:4-Qubits-QGHNN}) displays the results of the QGHNN model with four qubits.
	Heatmaps depict the Hamiltonian distribution of the target graph $G_t$, initial graph $G_i$, and learning result graph $G^*$. 
	The heatmaps in Fig. (\ref{fig:4-Qubits-QGHNN}) show that the Hamiltonian distribution of the learnt result graph $G^*$ approaches the target graph $G_t$ following system evolution via the QGHNN model.
	Table. (\ref{tab:Experiment01})  shows that the experimental accuracy of QGHNN can approach $99.8\%$ when employing four qubits to learn graph data, with a mean squared error of $0.004$. 
	When learning graph data with four qubits, QGHNN has the highest prediction accuracy and the most significant linear connection with the target graph. 
	Compared to the QAOA quantum machine learning algorithm, which also has advantages in graph computing, the computation results of the QGHNN model improve the accuracy by $8.6\%$ in cosine similarity and $9.2\%$ in correlation coefficient. 
	The Frobenius norm value decreases by $5.651$, while the MSE value decreases by $0.172$.
	Table. (\ref{tab:Experiment02}) illustrates the assessment metrics findings from Experiment (02), which was done through the PennyLane experimental quantum platform.
        Table. (\ref{tab:Experimental configuration02}) shows the parameter settings for Experiment (02).
	The assessment metrics in Table. (\ref{tab:Experiment02}) show that the cosine similarity of the graph $G^*$ acquired by QGHNN can reach $97.9\%$, with a mean squared error of $0.022$.
	When five qubits are used to learn graph data, QGHNN provides the highest accurate predictions as well as the best fitting effect on the target graph $G_t$.
	The noise impact in quantum circuits grows in proportion to the number of qubits used. 
	QGHNN learning outcomes in Experiment (02) are less accurate than in Experiment (01) due to the influence of quantum noise. 
	However, compared to other quantum machine learning models, QGHNN can retain high experimental accuracy, demonstrating its noise-resilient ability.
	Table. (\ref{tab:Experiment03}) highlights the assessment metrics findings from Experiment (03), which was done by the PennyLane experimental quantum platform. 
        Table. (\ref{tab:Experimental configuration03}) shows the parameter settings for Experiment (03).
	The assessment metrics in Table. (\ref{tab:Experiment03}) show that the experimental accuracy of QGHNN can reach $97.4\%$ when employing six qubits to learn graph data, with a mean squared error of $0.014$. 
	In Experiment (03), QGHNN outperforms all evaluation measures, with the lowest mean squared error, highest cosine similarity, lowest Frobenius norm, and highest correlation coefficient, which suggests that QGHNN is influential and trustworthy in high-dimensional quantum computing activities.
	
    Fig. (\ref{fig:loss function}) depicts the loss function of QGHNN in three different experiments with a training step size of $500$, illustrating that the loss function in each experiment follows a convergence trend. 
	In Experiment 01 (Fig. (\ref{fig:loss function}) (a)), the loss function begins at around $-0.993$, lowers steeply in the first stage with variations, and exhibits considerable modifications in the early training stage.
	The loss function stabilizes as training progresses and reaches -1.000 after around 300 steps. 
	The loss function patterns in Experiments (02) and (03) are comparable to those in Experiment (01), beginning at $-0.9825$ and $-0.986$, respectively, with considerable early decreases and following steady trends, eventually settling at around -1.000 after around $200$ to $300$ steps. 
    Fig. (\ref{fig:comparation02}) shows the results about the use of parameters between QGHNN and other different quantum machine learning models.
    QGHNN can achieve more accurate graph verification results by training the least parameters.
	
	Overall, the loss function in all three experiments shows significant early decreases, showing that the QGHNN model is actively optimizing and learning during this time.
	The oscillations in the loss function steadily decrease until they ultimately settle, indicating that the model parameters have achieved an optimal state and the loss value changes are reduced.
	Finally, the loss values in all experiments approach $-1.000$, demonstrating that the model has high convergence and stability under various experimental situations.
	In conclusion, the QGHNN model demonstrates adequate learning capacity and good convergence under various experimental settings, with the loss function finally stabilizing, suggesting that the model effectively reduced the loss and achieved the intended optimization effect.
	
	The comparative results in Table. (\ref{tab:Experiment01}), Table. (\ref{tab:Experiment02}), and Table. (\ref{tab:Experiment03}) show that regardless of whether the experiment has four qubits, five qubits, or six qubits, the QGHNN model outperforms all four assessment metrics.
	Mean squared error, cosine similarity, Frobenius norm, and correlation coefficient in QGHNN outperform other models across all qubit values, indicating that QGHNN has complete performance in quantum machine learning applications. 
	The mean squared error of QGHNN is lower for all qubit numbers, showing that QGHNN has vital prediction accuracy.
	However, when qubits grow, the mean squared error rises slightly from Table. (\ref{tab:Experiment01}) to Table. (\ref{tab:Experiment02}), which might be attributed to increasing training difficulty and data complexity as the model handles more qubits. 
	The cosine similarity of QGHNN stays high across varying qubit numbers, indicating a significant resemblance between learned graph $G^*$ and target graph $G_{t}$. 
	However, the cosine similarity declines slightly, demonstrating that the prediction result and the target graph become less similar as the number of qubits grows, which may imply a modest performance reduction of the model while dealing with higher-dimensional data.
	QGHNN has the lowest Frobenius norm among all qubit numbers, implying the least prediction error matrix.
	But the Frobenius norm steadily rises as the number of qubits grows, probably due to the cumulative effect of mistakes becoming more visible as the model processes increasingly complicated input. 
	The correlation coefficient of QGHNN stays high across various qubit values, demonstrating a solid linear link between model predictions and target data. 
	Although the correlation coefficient is reduced slightly, it remains the greatest, suggesting that QGHNN can capture the linear connection of different graphs well, even with high qubit numbers. 
	The high performance of QGHNN in all four assessment metrics may be attributed to its increased noise resilience. 
	When faced with noise, QGHNN can retain a lower mean squared error, a more remarkable cosine similarity, a lower Frobenius norm, and an increased correlation coefficient, which might be attributed to improved management of QGHNN of noise effects during model construction or training, making it more reliable in more complicated quantum computing jobs.

    \subsection{Advantages of QGHL and QGHNN}
    Presently, accessible quantum computers encounter significant challenges in hardware resource use. 
    Furthermore, when simulating QGHL and QGHNN using classical systems, the memory demands increase exponentially as the number of simulated qubits rises. 
    Consequently, QGHL is compared with the refs. \cite{ref.Wiebe2014HL}, \cite{ref.Paesani}, \cite{ref.HL.wang}, \cite{ref.shi2022parameterized.NEW}, to emphasize its importance in spontaneous system evolution, implementability on quantum computers, parameterized quantum circuits, graph learning capabilities, and noise resilience. 
    The detailed comparative outcomes are presented in Table. (\ref{tab:Advantages of QGHL}). 
    Table. (\ref{tab:Advantages of QGHNN}) presents a theoretical comparison of QGHNN with the refs. \cite{ref.VQE.du2022quantum}, \cite{ref.QAOA.guerreschi2019qaoa}, \cite{ref.zhao2024gqhan}, \cite{ref.bai2021learning}, \cite{ref.zhang2019quantum}, \cite{ref.dernbach2019quantum} emphasizing its advantages for parameterized quantum circuits, implementability on quantum computers, capability for graph representation, noise resilience, and methodologies for addressing graph learning.

    Table. (\ref{tab:Advantages of QGHL}) includes various key indications for QHL.
    First, both QGHL and  refs. \cite{ref.Wiebe2014HL}, \cite{ref.Paesani}, \cite{ref.shi2022parameterized.NEW} can actualize constantly evolving quantum systems, highlighting the beneficial function of QGHL in modeling the evolution of quantum Hamiltonian.
    The key difference is that QGHL can be completely implemented on a quantum computer, allowing the input and output of the system evolution process to be determined concurrently from its parameterized quantum circuit.
    In contrast, Wang's \cite{ref.HL.wang} and Shi's \cite{ref.shi2022parameterized.NEW} method includes extra conventional processing stages.
    Refs. \cite{ref.Wiebe2014HL}, \cite{ref.Paesani} apply quantum physics ideas to model the Hamiltonian of quantum systems without creating particular quantum circuits.
    QGHL incorporates numerous concepts from ref. \cite{ref.shi2022parameterized.NEW}, such as parameterized quantum circuit design and the system evolution process, and applies these concepts to create a unique operating mechanism, resulting in noise resistance of QGHL without precisely following the framework of ref.\cite{ref.shi2022parameterized.NEW}.
    Meanwhile, by defining the mapping connection between the Hamiltonian and the graph, QGHL can represent graph information.

    Table. (\ref{tab:Advantages of QGHNN}) lists numerous significant indications from the QML perspective.
    Initially, the methodologies of both QGHNN and Du's\cite{ref.VQE.du2022quantum}, Guerreschi's\cite{ref.QAOA.guerreschi2019qaoa}, Zhao's\cite{ref.zhao2024gqhan} are compatible with existing NISQ devices, highlighting the advantageous contribution of QGHNN to the domain of quantum machine learning. 
    The main difference is in the capability of QGHNN to completely execute graph learning representation on quantum processors while exhibiting noise immunity.
    On the other hand, the technique employed by Bai's\cite{ref.bai2021learning}, Zhang's\cite{ref.zhang2019quantum}, Dernbach's\cite{ref.dernbach2019quantum} employs quantum walk technology and incorporates supplementary classical simulation and processing stages.
    QGHNN incorporates several principles from Zhao's \cite{ref.zhao2024gqhan} methodology, such as employing gradient descent for function updates and the formulation of the loss function. 
    Meanwhile, QGHNN effectively employs the mapping connection between Hamiltonians and graphs to use contemporary NISQ devices for representing and learning graph data, mitigating the effects of quantum channel noise, and conserving classical storage.
	\begin{table*}[t]
		\centering
		\renewcommand\arraystretch{1.5}
		\renewcommand{\multirowsetup}{\centering}
		\caption{Advantages of QGHL}
		\label{tab:Advantages of QGHL}
		\setlength{\tabcolsep}{0.8mm}{
			\begin{tabular}{
			    p{6cm}<{\centering} p{2.2cm}<{\centering} p{2.2cm}<{\centering} p{2.2cm}<{\centering} p{2.2cm}<{\centering} p{2.2cm}<{\centering} }
				
				
                \hline
				\multicolumn{1}{c}{\multirow{2}{*}{\textbf{Indicators}}} 
				& \multicolumn{5}{c}{\textbf{Quantum Hamiltonian Learning Models}} 
                \\
                \cline{2-6}
				& Wiebe's\cite{ref.Wiebe2014HL}
				& Paesani's\cite{ref.Paesani}
				& Wang's\cite{ref.HL.wang}
				& Shi's\cite{ref.shi2022parameterized.NEW}
				& QGHL \\

				\hline
				\multicolumn{1}{c}{\multirow{1}{*}{Spontaneous System Evolution}} 
				& $\surd$
				& $\surd$
				& $\times$
				& $\surd$
				& $\surd$
			  \\
                \multicolumn{1}{c}{\multirow{1}{*}{Implementability on Quantum Computers}} 
				& $\times$
				& $\times$
				& Partial
				& Partial
				& Completely
				\\
                \multicolumn{1}{c}{\multirow{1}{*}{Parameterized Quantum Circuit }}
				& $-$
				& $-$
				& $\surd$
				& $\surd$
				& $\surd$
				\\
                \multicolumn{1}{c}{\multirow{1}{*}{Graph Learning Ability }} 
				& $-$
				& $-$
				& $-$
				& $-$
				& $\surd$
				\\
                \multicolumn{1}{c}{\multirow{1}{*}{Noise Resistant }} 
				& $\times$
				& $\times$
				& Weak
				& Weak
				& Excellent
				\\
                 \hline
		\end{tabular}}
	\end{table*}

    \begin{table*}[t]
		\centering
		\renewcommand\arraystretch{1.5}
		\renewcommand{\multirowsetup}{\centering}
		\caption{Advantages of QGHNN}
		\label{tab:Advantages of QGHNN}
		\setlength{\tabcolsep}{0.8mm}{
			\begin{tabular}{
			    p{2.0cm}<{\centering} p{1.5cm}<{\centering} p{2.0cm}<{\centering} p{1.5cm}<{\centering} p{1.5cm}<{\centering} p{2.0cm}<{\centering} p{2.0cm}<{\centering} p{1.5cm}<{\centering} }
				
				
                \hline
				\multicolumn{1}{c}{\multirow{2}{*}{\textbf{Indicators}}} 
				& \multicolumn{7}{c}{\textbf{Models}} 
                \\
                \cline{2-8}
				& Du's\cite{ref.VQE.du2022quantum}
				& Guerreschi's\cite{ref.QAOA.guerreschi2019qaoa}
				& Zhao's\cite{ref.zhao2024gqhan}
                & Bai's\cite{ref.bai2021learning}
                & Zhang's\cite{ref.zhang2019quantum}
				& Dernbach's\cite{ref.dernbach2019quantum}
				& QGHL \\

				\hline
				\multicolumn{1}{c}{\multirow{1}{*}{Implementability on Quantum Computers}} 
				& Partial
				& Partial
				& Partial
				& $\times$
				& $\times$
                & $\times$
                & Completely
			  \\
                \multicolumn{1}{c}{\multirow{1}{*}{Parameterized Quantum Circuit}} 
				& $\surd$
				& $\surd$
				& $\surd$
				& $-$
				& $-$
                & $-$
                & Excellent
				\\
                \multicolumn{1}{c}{\multirow{1}{*}{Noise Resistant}} 
				& Weak
				& Weak
				& Weak
				& $-$
				& $-$
                & $-$
                & Completely
				\\
                \multicolumn{1}{c}{\multirow{1}{*}{Methods for Solving Graph Learning}} 
				& VQE
				& QAOA
				& QML
				& Quantum Walk
				& Quantum Walk
                & Quantum Walk
                & QML
				\\
               
                \multicolumn{1}{c}{\multirow{1}{*}{Graph Representation Ability}} 
				& Weak
				& Well
				& Weak
				& Well
				& Well
                & Well
                & Excellent
				\\ 
                 \hline
		\end{tabular}}
	\end{table*}
	
	\section{Conclusion}
	QGHL is proposed in this work, where the implementation and the mapping correlation between graph $G$ and Hamiltonian $H_m$ on NISQ quantum devices are investigated. 
	QGHL shows how parameterized quantum circuits can be used to represent and learn graphs, which provides a basic and theoretical framework for future research into quantum machine learning that aims to solve non-Euclidean data issues.
	Furthermore, QGHNN is introduced to learn and represent graph $G$.
	The experiment results demonstrate that QGHNN exhibits both robustness and optimum performance across all assessment metrics, which achieves the lowest mean squared error of $0.004$ and the maximum cosine similarity of $99.8\%$. 
	Depending on current developments in quantum computing technology, QGHL, and QGHNN have promise for practical implementation in the future for building quantum knowledge graphs and quantum recommendation systems.

	
	\bibliographystyle{IEEEtran}
	\bibliography{ref}

\begin{thebibliography}{10}
\providecommand{\url}[1]{#1}
\csname url@samestyle\endcsname
\providecommand{\newblock}{\relax}
\providecommand{\bibinfo}[2]{#2}
\providecommand{\BIBentrySTDinterwordspacing}{\spaceskip=0pt\relax}
\providecommand{\BIBentryALTinterwordstretchfactor}{4}
\providecommand{\BIBentryALTinterwordspacing}{\spaceskip=\fontdimen2\font plus
\BIBentryALTinterwordstretchfactor\fontdimen3\font minus
  \fontdimen4\font\relax}
\providecommand{\BIBforeignlanguage}[2]{{%
\expandafter\ifx\csname l@#1\endcsname\relax
\typeout{** WARNING: IEEEtran.bst: No hyphenation pattern has been}%
\typeout{** loaded for the language `#1'. Using the pattern for}%
\typeout{** the default language instead.}%
\else
\language=\csname l@#1\endcsname
\fi
#2}}
\providecommand{\BIBdecl}{\relax}
\BIBdecl

\bibitem{ref.tetci.Dixit}
V.~Dixit, R.~Selvarajan, T.~Aldwairi, Y.~Koshka, M.~A. Novotny, T.~S. Humble,
  M.~A. Alam, and S.~Kais, ``Training a quantum annealing based restricted
  boltzmann machine on cybersecurity data,'' \emph{IEEE Transactions on
  Emerging Topics in Computational Intelligence}, vol.~6, no.~3, pp. 417--428,
  2022.

\bibitem{ref.tetci.Liu}
W.~Liu, Z.~Li, and Y.~Li, ``Quantum reachability games,'' \emph{IEEE
  Transactions on Emerging Topics in Computational Intelligence}, pp. 1--15,
  2024.

\bibitem{ref.huang2022quantum}
H.-Y. Huang, M.~Broughton, J.~Cotler, S.~Chen, J.~Li, M.~Mohseni, H.~Neven,
  R.~Babbush, R.~Kueng, J.~Preskill \emph{et~al.}, ``Quantum advantage in
  learning from experiments,'' \emph{Science}, vol. 376, no. 6598, pp.
  1182--1186, 2022.

\bibitem{ref.tcyb.04.shi}
J.~Shi, T.~Chen, W.~Lai, S.~Zhang, and X.~Li, ``Pretrained quantum-inspired
  deep neural network for natural language processing,'' \emph{IEEE
  Transactions on Cybernetics}, 2024.

\bibitem{ref.tetci.Yu}
H.~Yu and X.~Zhao, ``Event-based deep reinforcement learning for quantum
  control,'' \emph{IEEE Transactions on Emerging Topics in Computational
  Intelligence}, vol.~8, no.~1, pp. 548--562, 2024.

\bibitem{ref.zheng2024quantum}
J.~Zheng, Q.~Gao, M.~Ogorza{\l}ek, J.~L{\"u}, and Y.~Deng, ``A quantum spatial
  graph convolutional neural network model on quantum circuits,'' \emph{IEEE
  Transactions on Neural Networks and Learning Systems}, 2024.

\bibitem{ref.zhang.2022graph}
Y.~Zhang and Y.-M. Cheung, ``Graph-based dissimilarity measurement for cluster
  analysis of any-type-attributed data,'' \emph{IEEE Transactions on Neural
  Networks and Learning Systems}, vol.~34, no.~9, pp. 6530--6544, 2022.

\bibitem{ref.tetci.LiuGuanfeng}
G.~Liu, Y.~Wang, B.~Zheng, Z.~Li, and K.~Zheng, ``Strong social graph based
  trust-oriented graph pattern matching with multiple constraints,'' \emph{IEEE
  Transactions on Emerging Topics in Computational Intelligence}, vol.~4,
  no.~5, pp. 675--685, 2020.

\bibitem{ref.tetci.Zhu}
H.~Zhu, D.~Xu, Y.~Huang, Z.~Jin, W.~Ding, J.~Tong, and G.~Chong, ``Graph
  structure enhanced pre-training language model for knowledge graph
  completion,'' \emph{IEEE Transactions on Emerging Topics in Computational
  Intelligence}, vol.~8, no.~4, pp. 2697--2708, 2024.

\bibitem{ref.zhang.wang2021branch}
C.~Wang, Y.~Wang, and Y.~Cheung, ``A branch and bound irredundant graph
  algorithm for large-scale mlcs problems,'' \emph{Pattern Recognition}, vol.
  119, p. 108059, 2021.

\bibitem{ref.zhao2024gqhan}
R.-X. Zhao, J.~Shi, and X.~Li, ``Qksan: A quantum kernel self-attention
  network,'' \emph{IEEE Transactions on Pattern Analysis and Machine
  Intelligence}, 2024.

\bibitem{ref.QML.biamonte2017quantum}
J.~Biamonte, P.~Wittek, N.~Pancotti, P.~Rebentrost, N.~Wiebe, and S.~Lloyd,
  ``Quantum machine learning,'' \emph{Nature}, vol. 549, no. 7671, pp.
  195--202, 2017.

\bibitem{ref.bai2021learning}
L.~Bai, Y.~Jiao, L.~Cui, L.~Rossi, Y.~Wang, S.~Y. Philip, and E.~R. Hancock,
  ``Learning graph convolutional networks based on quantum vertex information
  propagation,'' \emph{IEEE Transactions on Knowledge and Data Engineering},
  vol.~35, no.~2, pp. 1747--1760, 2021.

\bibitem{ref.zhang2019quantum}
Z.~Zhang, D.~Chen, J.~Wang, L.~Bai, and E.~R. Hancock, ``Quantum-based subgraph
  convolutional neural networks,'' \emph{Pattern Recognition}, vol.~88, pp.
  38--49, 2019.

\bibitem{ref.dernbach2019quantum}
S.~Dernbach, A.~Mohseni-Kabir, S.~Pal, and D.~Towsley, ``Quantum walk neural
  networks for graph-structured data,'' in \emph{Complex Networks and Their
  Applications VII}, L.~M. Aiello, C.~Cherifi, H.~Cherifi, R.~Lambiotte,
  P.~Li{\'o}, and L.~M. Rocha, Eds.\hskip 1em plus 0.5em minus 0.4em\relax
  Cham: Springer International Publishing, 2019, pp. 182--193.

\bibitem{ref.Wiebe2014HL}
N.~Wiebe, C.~Granade, C.~Ferrie, and D.~Cory, ``Quantum hamiltonian learning
  using imperfect quantum resources,'' \emph{Physical Review A}, vol.~89,
  no.~4, p. 042314, 2014.

\bibitem{ref.hamiltonian.2024.NC}
A.~Gu, L.~Cincio, and P.~J. Coles, ``Practical hamiltonian learning with
  unitary dynamics and gibbs states,'' \emph{Nature Communications}, vol.~15,
  no.~1, p. 312, 2024.

\bibitem{ref.shi2022parameterized.NEW}
J.~Shi, W.~Wang, X.~Lou, S.~Zhang, and X.~Li, ``Parameterized hamiltonian
  learning with quantum circuit,'' \emph{IEEE Transactions on Pattern Analysis
  and Machine Intelligence}, vol.~45, no.~5, pp. 6086--6095, 2022.

\bibitem{ref.araz2023quantum.HL}
J.~Y. Araz and M.~Spannowsky, ``Quantum-probabilistic hamiltonian learning for
  generative modeling and anomaly detection,'' \emph{Physical Review A}, vol.
  108, no.~6, p. 062422, 2023.

\bibitem{ref.koch2023adversarial.HL}
R.~Koch, D.~Van~Driel, A.~Bordin, J.~L. Lado, and E.~Greplova, ``Adversarial
  hamiltonian learning of quantum dots in a minimal kitaev chain,''
  \emph{Physical Review A}, vol.~20, no.~4, p. 044081, 2023.

\bibitem{ref.quantum.gate.unitory.2021}
A.~W. Schlimgen, K.~Head-Marsden, L.~M. Sager, P.~Narang, and D.~A. Mazziotti,
  ``Quantum simulation of open quantum systems using a unitary decomposition of
  operators,'' \emph{Physical Review L}, vol. 127, no.~27, p. 270503, 2021.

\bibitem{ref.QHL.2023}
V.~Gebhart, R.~Santagati, A.~A. Gentile, E.~M. Gauger, D.~Craig, N.~Ares,
  L.~Banchi, F.~Marquardt, L.~Pezz{\`e}, and C.~Bonato, ``Learning quantum
  systems,'' \emph{Nature Reviews Physics}, vol.~5, no.~3, pp. 141--156, 2023.

\bibitem{ref.zhang.peng2022relation}
S.-J. Peng, Y.~He, X.~Liu, Y.-m. Cheung, X.~Xu, and Z.~Cui,
  ``Relation-aggregated cross-graph correlation learning for fine-grained
  image--text retrieval,'' \emph{IEEE Transactions on Neural Networks and
  Learning Systems}, vol.~35, no.~2, pp. 2194--2207, 2022.

\bibitem{ref.tetci.graph01}
L.~Huang, J.~Yuan, S.~Chen, and X.~Li, ``Mdg: A multi-task dynamic graph
  generation framework for multivariate time series forecasting,'' \emph{IEEE
  Transactions on Emerging Topics in Computational Intelligence}, vol.~8,
  no.~2, pp. 1337--1349, 2024.

\bibitem{ref.tetci.graph02}
G.~He, Z.~Zhang, H.~Wu, S.~Luo, and Y.~Liu, ``Kgcna: Knowledge graph
  collaborative neighbor awareness network for recommendation,'' \emph{IEEE
  Transactions on Emerging Topics in Computational Intelligence}, vol.~8,
  no.~4, pp. 2736--2748, 2024.

\bibitem{ref.tcyb.21.graph.LXL}
Q.~Wang, R.~Liu, M.~Chen, and X.~Li, ``Robust rank-constrained sparse learning:
  A graph-based framework for single view and multiview clustering,''
  \emph{IEEE Transactions on Cybernetics}, vol.~52, no.~10, pp.
  10\,228--10\,239, 2021.

\bibitem{ref.zhang.liu2022learning}
X.~Liu, Y.~He, Y.-M. Cheung, X.~Xu, and N.~Wang, ``Learning
  relationship-enhanced semantic graph for fine-grained image--text matching,''
  \emph{IEEE Transactions on Cybernetics}, vol.~54, no.~2, pp. 948--961, 2022.

\bibitem{ref.tetic.LXL.zhang2020reconstructing}
Y.~Zhang, C.~Yang, K.~Huang, M.~Jusup, Z.~Wang, and X.~Li, ``Reconstructing
  heterogeneous networks via compressive sensing and clustering,'' \emph{IEEE
  Transactions on Emerging Topics in Computational Intelligence}, vol.~5,
  no.~6, pp. 920--930, 2020.

\bibitem{ref.tetic.zhang.peng2022relation}
S.-J. Peng, Y.~He, X.~Liu, Y.-m. Cheung, X.~Xu, and Z.~Cui,
  ``Relation-aggregated cross-graph correlation learning for fine-grained
  image--text retrieval,'' \emph{IEEE Transactions on Neural Networks and
  Learning Systems}, vol.~35, no.~2, pp. 2194--2207, 2022.

\bibitem{ref.zhang.yang2023sagn}
Y.~Yang, X.~Tang, Y.-M. Cheung, X.~Zhang, and L.~Jiao, ``Sagn: Semantic-aware
  graph network for remote sensing scene classification,'' \emph{IEEE
  Transactions on Image Processing}, vol.~32, pp. 1011--1025, 2023.

\bibitem{ref.tetic.LXL.liu2024reinforcement}
Y.~Liu, Y.~Pang, R.~Jin, Y.~Hou, and X.~Li, ``Reinforcement learning and
  transformer for fast magnetic resonance imaging scan,'' \emph{IEEE
  Transactions on Emerging Topics in Computational Intelligence}, 2024.

\bibitem{ref.LXL.zhang2024decouple}
H.~Zhang, Y.~Zhu, and X.~Li, ``Decouple graph neural networks: Train multiple
  simple gnns simultaneously instead of one,'' \emph{IEEE Transactions on
  Pattern Analysis and Machine Intelligence}, 2024.

\bibitem{ref.zhang.chen2024qgrl}
J.~Chen, Y.~Ji, R.~Zou, Y.~Zhang, and Y.-m. Cheung, ``Qgrl: Quaternion graph
  representation learning for heterogeneous feature data clustering,'' in
  \emph{Proceedings of the 30th ACM SIGKDD Conference on Knowledge Discovery
  and Data Mining}, 2024, pp. 297--306.

\bibitem{ref.VQE.du2022quantum}
Y.~Du, T.~Huang, S.~You, M.-H. Hsieh, and D.~Tao, ``Quantum circuit
  architecture search for variational quantum algorithms,'' \emph{npj Quantum
  Information}, vol.~8, no.~1, p.~62, 2022.

\bibitem{ref.QAOA.guerreschi2019qaoa}
G.~G. Guerreschi and A.~Y. Matsuura, ``Qaoa for max-cut requires hundreds of
  qubits for quantum speed-up,'' \emph{Scientific reports}, vol.~9, no.~1, p.
  6903, 2019.

\bibitem{ref.Paesani}
S.~Paesani, J.~Wang, R.~Santagati, S.~Knauer, A.~A. Gentile, N.~Wiebe,
  M.~Petruzzella, A.~Laing, J.~G. Rarity, J.~L. O’Brien \emph{et~al.},
  ``Experimental quantum hamiltonian learning using a silicon photonic chip and
  a nitrogen-vacancy electron spin in diamond,'' in \emph{European Quantum
  Electronics Conference}.\hskip 1em plus 0.5em minus 0.4em\relax Optical
  Society of America, 2017.

\bibitem{ref.HL.wang}
Y.~Wang, G.~Li, and X.~Wang, ``A hybrid quantum-classical hamiltonian learning
  algorithm,'' \emph{Science China-Information Sciences}, vol.~66, no.~2, 2023.

\end{thebibliography}

	

	\vfill
	
\end{document}